\documentclass[11pt,preprint]{aastex}  
\usepackage{epsfig}             

\newcommand{\eten}[1]{\mbox{$10^{#1}$}}
\newcommand{\ee}[1]{\mbox{${} \times 10^{#1}$}}
\newcommand{\msun}{\mbox{$M_\odot$}}
\newcommand{\lsun}{\mbox{$L_\odot$}}

\newcommand{\tff}{\mbox{$t_{\rm{ff}}$}}

\newcommand{\tmin}{\mbox{$T_{\rm min}$}}
\newcommand{\td}{\mbox{$T_{\rm dust}$}}
\newcommand{\tg}{\mbox{$T_{\rm gas}$}}
\newcommand{\tk}{\mbox{$T_{\rm K}$}}
\newcommand{\tkin}{\mbox{$T_{\rm K}$}}
\newcommand{\kms}{\mbox{km s$^{-1}$}}
\newcommand{\sfrff}{\mbox{$\rm SFR_{ff}$}}
\newcommand{\cmv}{\mbox{cm$^{-3}$}}
\newcommand{\lfir}{\mbox{$L_{\rm FIR}$}}

\shorttitle{Dust and Gas Energetics}
\shortauthors{Urban et al.}

\begin{document}

\title{Fragmentation and Evolution of Molecular Clouds. III: The Effect of Dust
and Gas Energetics}

\author{Hugo Martel\altaffilmark{1,2}, Andrea Urban\altaffilmark{3}, 
and Neal J. Evans II\altaffilmark{4}}

\altaffiltext{1}{D\'epartement de physique, de g\'enie physique et d'optique,
Universit\'e Laval, Qu\'ebec, QC, G1V 0A6, Canada}

\altaffiltext{2}{Centre de Recherche en Astrophysique du Qu\'ebec}

\altaffiltext{3}{
Sapling Learning, Inc.
}

\altaffiltext{4}{
Department of Astronomy, University of Texas,
2515 Speedway, Stop C1400,
Austin, TX 78712-1205,
USA
}

\begin{abstract}
Dust and gas energetics are incorporated into a cluster-scale simulation of
star formation in order to study the effect of heating and cooling on the star
formation process.  We build on our previous work by calculating separately the
dust and gas temperatures.  The dust temperature is set by radiative equilibrium
between heating by embedded stars and radiation from dust.  The gas temperature
is determined using an energy-rate balance algorithm which includes molecular
cooling, dust-gas collisional energy transfer, and cosmic-ray ionization.
The fragmentation proceeds roughly similarly to simulations in which the gas
temperature is set to the dust temperature, but there are differences.
The structure of regions around sink particles have properties similar to 
those of Class 0 objects, but the infall speeds and mass accretion rates
were, on average, higher than those seen for regions forming only low-mass
stars.
The gas and dust temperature have complex distributions not
well modeled by approximations that ignore the detailed thermal physics.
There is no simple relationship between density and kinetic temperature.
In particular, high density regions have a large range of temperatures,
determined by their location relative to heating sources.
The total luminosity underestimates the star formation rate
at these early stages, before ionizing sources are included,  
by an order of magnitude.
As predicted in our previous work, a larger number of intermediate mass
objects form when improved thermal physics is included, 
but the resulting IMF still has too few low mass stars.
However, if we consider recent evidence on core-to-star efficiencies, the
match to the IMF is improved.
\end{abstract}


\section{Introduction}

The physics of star formation is the link between the small-scale -- 
planet formation -- and the large-scale -- galactic evolution. 
Many, probably most, stars form in highly clustered environments
(\citealt{lada03,bressert10}).
Within the highly complex structure of a molecular cloud, theorists
have identified the clump as the object that forms a cluster
(\citealt{williams2000,mckee07}).
Observational studies have found a range of structures that might
correspond to this concept, but the formation of rich clusters
seems most clearly associated with particularly dense clumps, identified
by strong emission from tracers of dense gas 
(e.g., \citealt{wu2010}).

This paper is the third of a series that studies the fragmentation of
a dense molecular clump using Smoothed Particle Hydrodynamics (SPH)
to follow the hydrodynamics, with a focus on 
the effects of the thermal physics. In Paper~I
(\citealt{mes06}, hereafter MES06), we showed that 
an isothermal gas will fragment excessively, producing only very low
mass stars.
In Paper II (\citealt{Urban10}, hereafter UME10),
we included global radiative feedback from the forming
stars, assuming that the gas temperature was equal to the dust 
temperature; this approximation over-produced high mass stars.
In this paper, we calculate the gas temperature separately from the
dust temperature.
We use these simulations to address the role of thermal physics in the
fragmentation problem, 
the density distribution around forming stars in a proto-cluster, 
the evolution of the far-infrared luminosity during formation, with
application to the use of far-infrared luminosity as a probe of
star formation rate, and the effects of an improved thermal physics
on the mass function of forming stars. 

In our previous work (UME10) we modeled a clustered star-forming region and
showed that dust-gas thermal energetics with source luminosity terms from young
stars can heat the gas and prevent fragmentation. Less fragmentation led to
the formation of massive stars, which had not been produced in our isothermal
simulation (MES06).
Although we were able to form massive stars, we
missed a significant fraction of the low-mass stellar population.  We
hypothesized in UME10 that by including a more realistic dust-gas thermal
energetics algorithm we would increase the number of low-mass stars.  Including
molecular cooling would decrease the temperature of the gas, leading to more
fragmentation and more low-mass stars. To test our hypothesis, we have
implemented the complete heating and cooling algorithm described in
\citet{UrbanDG} (hereafter UED09)
in simulations with similar scales and parameters as the
simulations discussed in UME10. We discuss our work in the
following sections. In \S\ref{sec:code}, we discuss our numerical algorithm
and our new method of calculating the gas temperature. In \S\ref{sec:sim}, we
discuss the initial conditions and parameters of our simulation.  Our results
are discussed in \S\ref{sec:results}. We conclude and summarize
the paper in \S\ref{sec:con}.

\section{The Numerical Algorithm}\label{sec:code}

Our numerical algorithm was described in MES06 and UME10. It is a standard 
Smoothed Particle Hydrodynamics (SPH) algorithm (see \citealt{monaghan},
references therein), which simulates the growth of structures in a 
cubic volume with periodic boundary conditions, representing a small
part of a giant molecular cloud. 
The code was modified to include particle splitting and
sink particles. In the optically-thin regime, the Jeans mass
decreases with increasing density. Eventually, when the density becomes
sufficiently high, the gas becomes optically thick
and the Jeans mass starts increasing with density.
Hence, there is a minimum Jeans mass, corresponding to the
transition between the two regimes. To properly follow the fragmentation
of the cloud, it is essential to resolve that minimum Jeans mass.
Particle splitting (see \citealt{kw}; MES06) enables us
to do this at a reasonable computational cost.

When the gas reaches a certain critical density $\rho_c$, 
the algorithm replaces
gas fragments by {\it sink particles}, using the method of
\citet{bcl}. Sink particles (or sinks) represent protostellar cores.
They are not allowed to fragment or merge, but they have the
ability to grow by accreting surrounding gas particles. 
Any bound gas particle within its accretion radius, $r_{\rm acc}$ 
($\sim150\,{\rm AU}$ for all but one simulation, see Table \ref{tbl:params}),
is automatically accreted into the sink.  
Because $r_{\rm acc}$  is considerably larger than the actual forming star
(likely radius a few to tens of solar radii),
the evolution of material inside the sink is unknown. For the simulations,
we assume that all the material falling into the sink flows continuously
onto the actual stellar core, producing accretion luminosity.
The luminosity resulting from that accretion will heat the surrounding gas.
In order to calculate the luminosity from a sink particle, we use the models of
\citet{wt}, specifically their Table 3, which include the effect of mass
accretion on the luminosity.  For objects with masses greater than $2\msun$, we
use the method described in UME10 to calculate the luminosity.
There is considerable evidence that not all material passing through
a radius of $150\,{\rm AU}$ winds up in the forming star, so we also consider
the effects of the loss of some material in comparing to the mass
function (\S\ref{sec:MF}).

\subsection {Dust and Gas Temperature Calculation}\label{sec:tdust}

We use the same numerical methods described in UME10  to calculate the mass
accretion rate onto sinks  and the dust temperature.  In order to calculate the
gas temperature, we use the method of UED09.  
We give a brief description here.

The dust temperature is determined using the method discussed in UME10, and
in more detail in UED09. 
UED09 used a spherically-symmetric radiative transfer code, DUSTY
\citep{dusty}, to calculate the dust temperature distribution around young
stellar objects.  
Using DUSTY, we created a grid of models with input values
of luminosity and density distribution.  In the simulations presented in this
paper, we calculate the  luminosity of individual objects based on mass and
mass accretion rate using the models of \citet{wt}.  We also determine the
density profile around each of the sinks formed in our simulation.  We then use
the grid calculated in UED09 to find an analytic fit to the dust
temperature distribution around each of the individual sinks.  We approximate
the dust/gas distribution around sink particles as spherical.

We calculate the density profile around individual sink particles, as in
UME10, using spherical shells. We parameterize the density profile with $n_o$
and $\alpha$, as the following,
\begin{equation}\label{eq:n}
n(r)  = n_o \left(\frac{r}{1000\,\textrm{AU}} \right)^{-\alpha}
\textrm{cm}^{-3}.
\end{equation}

\noindent
The symbol $n$ represents the number density of all particles 
($n=n_{\rm H_2}^{\phantom1}+n_{\rm He}^{\phantom1}$).
We assume a ratio $n_{\rm{H}_2}/n_{\rm{He}}=5$,
which corresponds to a mean molecular weight
$\mu=2.33$. The gas density is $\rho=\mu m_{\rm H}^{\phantom1}n$.

Our approximations are complementary to those of most other simulations 
that include some thermal physics. 
We do not include compressional heating during collapse, as does \cite{bate09r}.
We are not doing radiative transfer during the SPH calculation, but
instead using a pre-computed grid as did \citet{smith}, 
and we are assuming spherical distributions of material around sinks. 
Thus, we will miss some of the effects of geometry included in papers 
that do not assume spherical symmetry
(e.g., \citealt{krumholz}, \citealt{krumholz10}, \citealt{bate09r}, 
\citealt{offner}).
On the other hand, we used realistic grain opacities as a function of
wavelength (i.e., OH5 dust opacities, \citealt{oh5}, as described 
in \citealt{chad}) in radiative transfer calculations that
include non-isotropic scattering and apply to all relevant optical depths, 
in contrast to mean opacities and other
approximations used by most other simulations 
(e.g., \citealt{krumholz, krumholz12}).

The gas temperature algorithm (UDE09)
includes energy transfer between gas and dust
via collisions, gas heating by cosmic rays, and molecular cooling.
Heating by photoelectric emission from dust grains is not included
because the external ultraviolet radiation is strongly attenuated and the
clump we are simulating is assumed to reside deep in a larger molecular
cloud. We are not including ultraviolet radiation from the forming stars.
The dust to gas ratio is taken to be 4.86\ee{-3} 
\citep{1989ApJ...342..306H},
and the grain cross section per baryon is 6.09\ee{-22} cm$^2$
\citep{2004ApJ...614..252Y}.
The cosmic ray ionization rate is 3.0\ee{-17} s$^{-1}$ 
\citep{2000A&A...358L..79V}
and the energy deposited per ionization is 20 eV \citep{2001ApJ...557..736G}. 
The fractional abundance of CO is taken to be \eten{-4}.
The algorithm requires inputs of dust temperature (discussed in
the previous paragraph), local velocity dispersion ($\delta v$), 
column density, and
local density. We use the local density calculated from the density profile
[eq.~(\ref{eq:n})]
to be self-consistent with our dust temperature calculation.  The local
velocity and the column density are needed in order to calculate the level of
radiative trapping in the molecular cooling lines. The local velocity 
dispersion
(characterized by the Doppler $b$ parameter) was
assumed to be $1\,\kms$  throughout the calculation.  This is a reasonable
assumption based on the values of the velocity field found in UME10 
(see Table~\ref{tbl:lmae} below).

We estimate the column density at a point of interest 
${\bf r}$ within our simulation with the following line integral, for
every sink $i$ present in the simulation at that time:
\begin{equation}\label{eq:ncol}
N_{{\rm column},i} = \int^{\Delta l_i+2000{\rm AU}}_{\Delta l_i} n_i(r)d{\bf\hat r},
\end{equation}

\noindent
where $\Delta l_i$ is the distance from sink $i$ to ${\bf r}$,
$n_i(r)$ is the density profile around sink $i$ calculated using 
equation~(\ref{eq:n}), ${d{\bf\hat{r}}}$ indicates 
that the direction of integration is radial. Hence, we are essentially 
calculating the column density of gas along a line of length $2000\,{\rm AU}$
starting at the point of interest and pointing away from the sink.
The sink particle that gives the
highest column density at the point of interest is used to calculate the
column density and also the local density at that point within the simulation.
The local density is calculated using the density profile of the chosen sink.
The choice of highest column density is made to ensure that we are not
calculating the column density from the tail of the distribution around
a more distant sink particle (see UED09).
The limit of integration was set at $2000\,{\rm AU}$ to agree with 
the value used in UED09.  

As in UME10, we impose a minimum temperature $\tmin$ to the gas. Hence,
if the calculation of $\tg$ produces a value lower than $\tmin$, we
use $\tg=\tmin$ instead.
We consider the effects of changing the value of \tmin.

\section{The Simulations}\label{sec:sim}

\begin{deluxetable}{lccrcccc}
\tablecaption{Numerical Parameters of the Simulations}
\tablewidth{0pt}
\tablehead{
\colhead{Run} & \colhead{$\tmin[{\rm K}]$} &
\colhead{$\tk$} &
\colhead{$M_{\rm tot}$ $[\msun]$} &
\colhead{$L_{\rm box}$ $[{\rm pc}]$} &
\colhead{$M_J^{\rm init}$ $[\msun]$} & 
\colhead{$M_J$ $[\msun]$} &
\colhead{$r_{\rm acc}$[AU]}
}
\startdata
 I05 &  5 & $\tmin$ &  671.4 & 0.984 & 0.617 & 0.0080 & 152 \cr
 D05 &  5 & $\td$   &  671.4 & 0.984 & 0.617 & 0.0080 & 152 \cr
 G05 &  5 & $\tg$   &  671.4 & 0.984 & 0.617 & 0.0080 & 152 \cr
 G10 & 10 & $\tg$   & 1898.0 & 1.390 & 1.744 & 0.0226 & 215 \cr
\enddata
\label{tbl:params}
\end{deluxetable}

We performed two new simulations, G05 and G10, which include all the 
dust energetics discussed in \S\ref{sec:tdust}, but also 
add a more accurate treatment of the gas energetics 
described in more detail in UED09. 
In the following results section, we will refer to these new simulations
as using a ``complete energetics'' algorithm.
For comparison, we
also include two simulations, I05 and D05, that were presented in UME10.
Our initial conditions are identical to those
described in UME10.
The parameters of the simulations are given in Table~\ref{tbl:params}.
In the third column, $\td$ and $\tg$ refer to the dust
and gas temperature calculated using the complete energetics algorithm,
while $\tk$ is the actual temperature that we use for the gas. 
The letters I, D, and G stand for ``isothermal,'' ``dust,'' and
``gas,'' respectively.

In all simulations, the initial density of our cloud is 
$\bar\rho=4.75\times10^{-20}{\rm g\,cm^{-3}}$,
or $\bar n=1.22\times10^4{\rm cm}^{-3}$
assuming $\mu = 2.33$. 
This density is similar to the median average density 
($\bar n = 1.6\ee4$) of a well-studied
sample of massive dense clumps \citep{wu2010}.
In simulations I05, D05, and G05, the minimum
temperature of the gas is set at $\tmin=5{\rm K}$, which corresponds to
an initial Jeans mass $M_J^{\rm init}=0.617\msun$. 
Sink particles are created at a threshold density of
$\rho_c=2.822\times10^{-16}\,{\rm g\,cm^{-3}}$, 
or $n_c=7.252\times10^{7}{\rm cm}^{-3}$, which represents a contraction
by a factor of 5942. 
The corresponding Jeans
mass at $\rho=\rho_c$ is $M_J=0.008\msun$. As in UME10, we adjust the particle
mass such that each Jeans mass is resolved with 200 particles. The simulation
starts with $64^3$ particles, but we allow 2 levels of particle splitting
($N_{\rm gen}=2$, see UME10). The effective number of particles is
therefore $256^3$, the total mass of the system is
$M_{\rm tot}=(256^3/200)M_J=671.4\msun$, and the box size is
$L_{\rm box}=(M_{\rm tot}/\bar\rho)^{1/3}=0.984\,{\rm pc}$.
Our SPH code uses a standard cubic spline smoothing kernel. The individual
smoothing lengths are adjusted dynamically 
so that each gas particle has about 50 neighbors. 

All simulations use identical initial conditions. Particles are laid down
on a $64^3$ cubic grid, and displaced in order to reproduce a Gaussian
random density fluctuation with a $P(k)\propto k^{-2}$ power spectrum
(\citealt{klessen}, MES06). 
Initial velocities are then adjusted in order to reproduce
a pure growing mode (MES06).
In simulation G05, we set the minimum temperature to $5\,{\rm K}$ to allow a
direct comparison with simulations I05 and D05 taken from UME10.
However, most observations suggest $\tk\geq10\,{\rm K}$, except in 
well-shielded areas of dense cores. Although massive, 
dense clumps may in fact be quite cold before they form stars, we explore the
effects of $\tmin$ with a second simulation with full energetics,
G10, in which we used $\tmin=10\,{\rm K}$. We did not change
the initial density $\bar\rho$ and the threshold density $\rho_c$.
Doubling the temperature while keeping $\rho_c$ fixed increases
the minimum Jeans mass by a factor of $2^{3/2}$, up to $0.0226\msun$.
Following the approach used in UME10, we decided to keep the same 
resolution for all simulations: 200 particles per Jeans mass. As a result,
the particle mass and total mass increase by a factor or $2^{3/2}$, and
the box size increases by a factor of $2^{1/2}$. The total mass
$M_{\rm tot}$, box size $L_{\rm box}$, the initial Jeans mass
$M_J^{\rm init}$, Jeans mass $M_J$ at sink formation, and
accretion radius $r_{\rm acc}$ are listed in Table~\ref{tbl:params}.

In all simulations, the first particle splitting occurs at a density
$\rho=5.80\bar\rho$, the second one at density $\rho=371\bar\rho$, and
sink formation starts at density $\rho=\rho_c=5942\bar\rho$. The
first sinks formed have an initial mass $M\approx M_J$. Sinks formed
afterward will tend to be initially more massive since gas heating by the
first sinks increases the Jeans mass.
The initial free-fall time in our simulations is 
$\tff= 9.64\ee{12}{\rm s}=3.06\ee{5}{\rm yr}$.
We run our simulations for a few free-fall times until the most
massive sink particle in the simulation reaches $M\approx21\msun$. We halt
the simulations at this point since the luminosity from massive stars will
produce significant ionizing photons.  These photons will then dominate the
evolution of the simulation as seen in \citet{Dale2005}.

\section{Results}\label{sec:results}

\subsection{Fragmentation and Sink Formation}

\begin{figure}
\begin{center}
\includegraphics[width=6in]{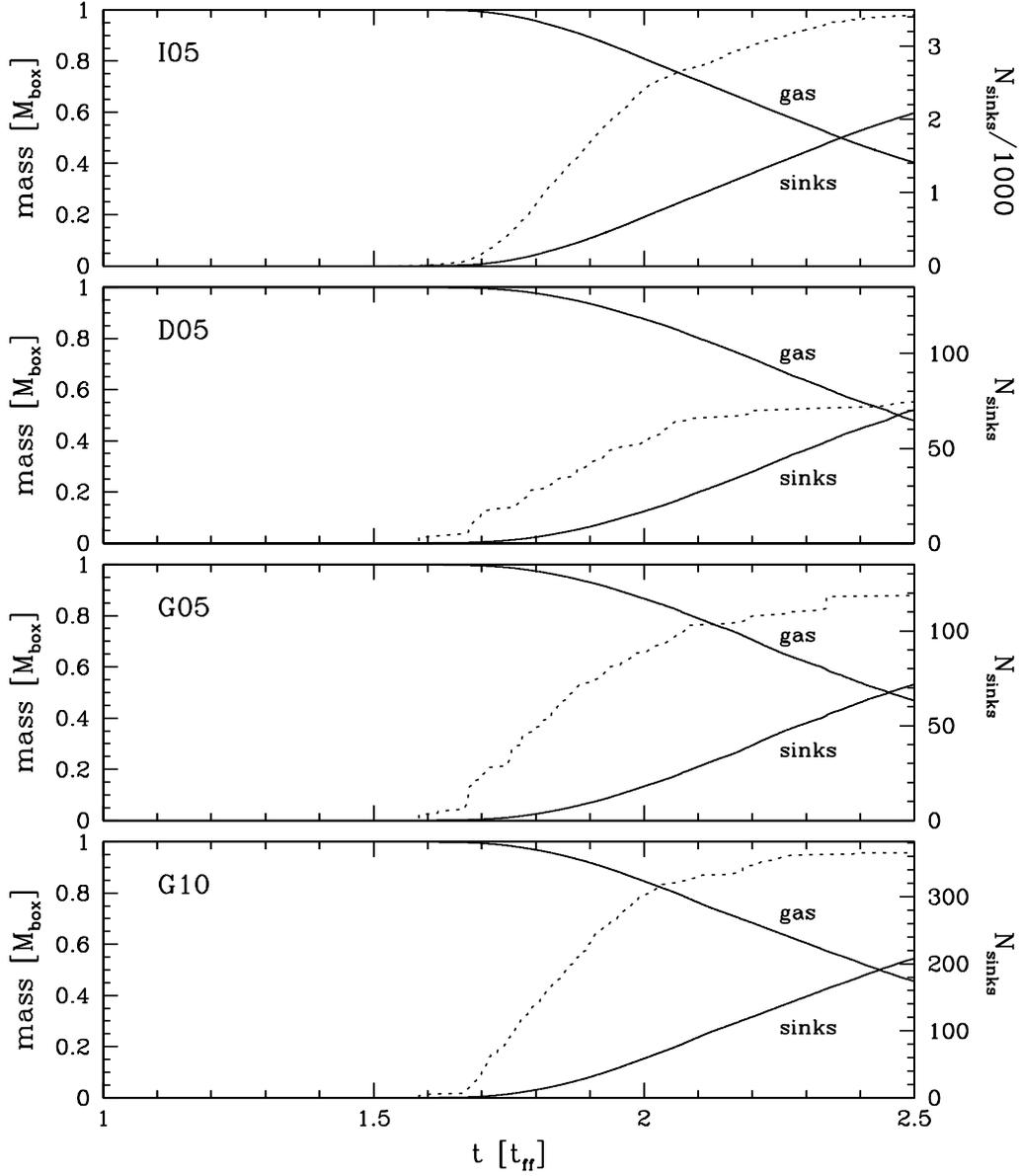}
\caption{Evolution of the mass fraction in gas and sinks, and
the number of sinks, for all four simulations.
Top panel: run I05; second panel: run D05; third panel: run G05;
bottom panel: run G10.
Top solid lines show the mass fraction in gas.  
Bottom solid lines show the mass fraction in sinks.
Dotted lines with the scales on the right axes show the number of sinks. 
The scales of the right axes in the first and fourth panels are different
from those in the second and third panel. The one in the fourth
panel is larger than the one in the third panel
by a factor of $2^{3/2}$ to account for the
increased volume.
}
\label{fig:gfrac1}
\end{center}
\end{figure}

Figure \ref{fig:gfrac1} shows the fractional mass
distribution of the gas and sink particles as a function of time for the
various simulations. 
The mass fraction in gas and in sinks is nearly identical for all runs
with feedback (D05, G05, and G10). In all cases, the transition from
gas- to sink-dominated mass fraction 
takes place around $t=2.45\tff$, while for run
I05 this transition takes place earlier, at $t=2.36\tff$. By 
$t=2.4\tff$, we have formed 74 sinks in run D05 and 118 in run G05.
The slower formation of sink particles in run D05, for which 
$\tk=\td$ is due to the fact that the dust temperature
is usually larger than the gas temperature, and setting the gas
temperature to that higher value tends to
inhibit the formation of sinks. We discuss this in more
detail in \S\ref{sec:tempdens}.

It is interesting to compare runs G05 and G10. As we explained in \S3,
the volume simulated in run G10 is larger than the one simulated in run G05
by a factor of $2^{3/2}$. Hence, because run G05 formed 118 sinks by 
$t=2.4\tff$, we would expect run G10 to form about 334 sinks just
because of its increased mass, not accounting for the effect of temperature. 
Run G10 actually forms 365 sinks by $t=2.4\tff$,
within 8\% of our prediction. The right axis in the bottom panel of 
Figure~\ref{fig:gfrac1} has been rescaled by a factor of $2^{3/2}$ compared
to the third panel to allow a direct visual comparison. Sinks form faster
in run G10 than run G05, but eventually the number of sinks 
{\it per unit volume} becomes comparable. Changing the minimum temperature
affects the formation of the first sinks. But once several sinks have formed,
the minimum temperature becomes irrelevant, because feedback heating by
sinks raises the gas temperature above 10K in regions where the next sinks
form, as we show in \S\ref{sec:tempdens} below.

\begin{deluxetable}{ccccccc}
\tablecaption{Final State for all Simulations}
\tablewidth{0pt}
\tablehead{
\colhead {Run} & 
\colhead{$t_{\rm final}$ $[\tff]$} & \colhead{$N_{\rm sinks, final}$}
& \colhead {$M_{\rm sink,max}$ $[\msun]$}
& \colhead {$f_{\rm sinks,final}$} &\colhead{SFR$_{\rm ff}$}
}
\startdata
I05 & 2.5 & 3429 & 0.50 & 60\% & 0.24\cr
D05 & 2.5 &   74 & 20.8 & 50\% & 0.20\cr
G05 & 2.4 &  118 & 20.8 & 46\% & 0.19\cr
G10 & 2.4 &  365 & 24.0 & 47\% & 0.20\cr
\enddata
\label{tbl:sim}
\end{deluxetable}

In Table \ref{tbl:sim}, we compare the final state of all four simulations.
The third, fourth, and fifth column  give the number of sinks, maximum 
sink mass, and mass fraction in sinks at the final time, respectively.
Comparing runs D05 and G05, which both have a total mass of $671\msun$,
we find that fewer sinks are formed in run D05, but the gas that
is prevented from forming new sinks ends up accreting onto the existing sinks,
so the total mass in sinks is roughly the same at the end of the simulations.

\begin{figure}
\epsscale {1.0}
\includegraphics[width=6in]{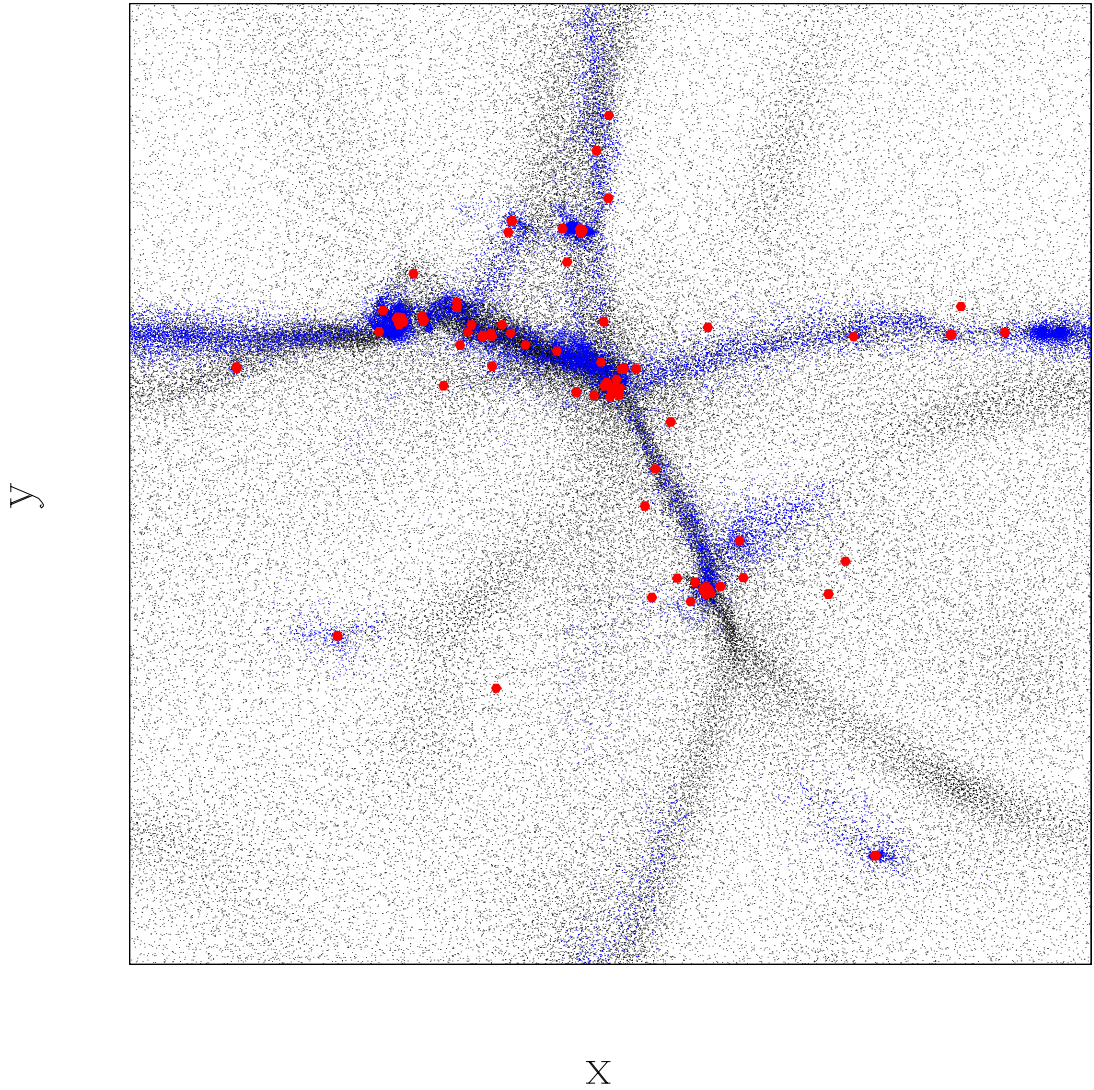}
\caption{XY position plot of sinks and gas particles
at $t=2.4\tff$ for run G05. 
Black and blue dots indicate gas particles.
Blue dots are gas particles which have undergone one particle splitting.
Red dots are sinks. The box is 0.984 pc $\times$ 0.984 pc.
}
\label{fig:gxy2}	
\end{figure}

\begin{figure}
\begin{center}
\includegraphics[width=6in]{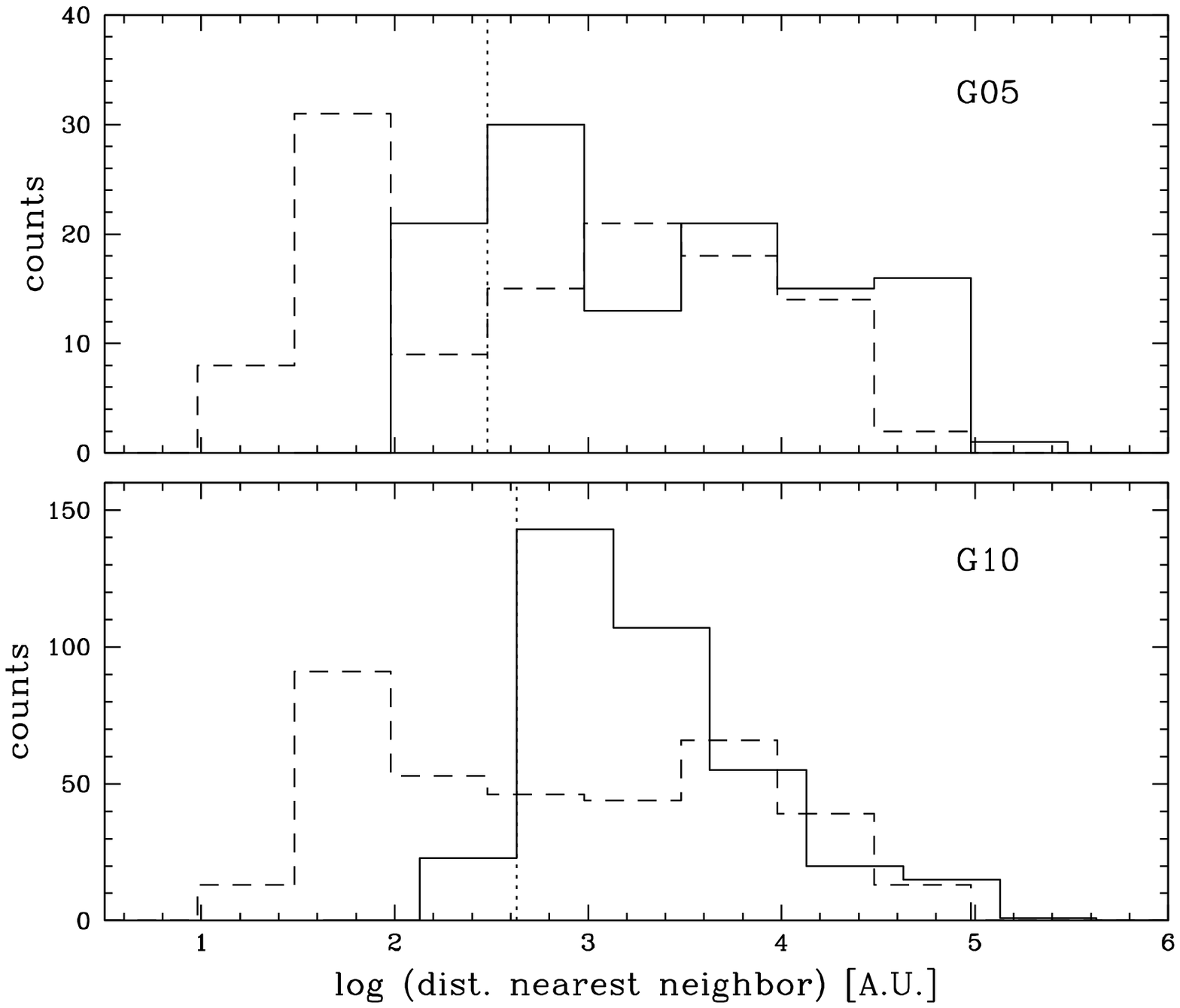}
\caption{Solid histograms: separations between each sink at the time of
its formation and the nearest already existing sink at that time.
Dashed histograms: separations between each sink and its
nearest neighbor at the end of the simulation.
Top panel: run G05; bottom panel: run G10.
Dotted lines indicate the value of $2r_{\rm acc}$. }
\label{fig:histo2}
\end{center}
\end{figure}

Figure \ref{fig:gxy2}
shows the distribution of sink and gas particles in our simulation box at 
$t=2.4\tff$ for run G05. 
We chose this time because this is when a substantial fraction of
the total number of final sinks have formed and there is also a significant
amount of gas remaining, as seen in Figure~\ref{fig:gfrac1}.
The morphology of the gas shown in this figure is similar
to that seen in similar figures in UME10. There are fewer sinks formed
compared to the isothermal simulation I05, but there are more compared to the
simulation with only dust heating D05.
The distribution of sink particles near the
intersection of filaments is similar to what is seen in \citet{bate09r}.

The sinks form in close proximity. To quantify this, starting with sink~\#2,
every time a sink formed, we calculated the distance to the nearest existing
sink. The resulting distributions are shown as solid
histograms in Figure~\ref{fig:histo2}.
There is a ``natural'' separation $\Delta r=2r_{\rm acc}$, which comes from
the method used for forming sinks. The algorithm converts gaseous spheres of
radius $r_{\rm acc}$ into sinks. When a dense region fragments into several
sinks, the minimum separation is of order $2r_{\rm acc}$, 
corresponding to spheres of gas that are in contact. 
Also, when a new sink forms near an existing one, the separation 
tends to be of order $2r_{\rm acc}$: the gas density profiles
near sinks (\S \ref{sec:ddpe}) show that the density decreases with 
increasing radius; hence the gas located
the closest to the existing sink will be the first to reach the threshold
density $\rho_c$. The large separations seen in Figure~\ref{fig:histo2}
correspond
to sinks that were the first to form in a gas clump that was located
away from all other clumps where sinks were already present.
Although there is a physical reason for sinks to form
close to one another, the particular value of $2r_{\rm acc}$ has no physical
meaning. The parameter $r_{\rm acc}$ is adjustable, 
and so is the density threshold
$\rho_c$. Only the combination $r_{\rm acc}^3\rho_c$ has a physical meaning,
since it determines the Jeans mass $M_J$. We could have used a larger
value of $r_{\rm acc}$ and a correspondingly smaller value of $\rho_c$.
Sinks would have formed farther apart, but gravity would have brought them
together, and the end result would have been essentially the same.
In very dense regions, separations between sinks can reach values as
low as $10-20\,{\rm AU}$. They are clearly much closer than they were
when they formed.

Although some sinks move closer together, others move apart.
The final distribution of separations is broad and skewed 
(dashed histograms in Fig. \ref{fig:histo2}), indeed bimodal. 
By the end of simulations G05 and G10,
the mean separation is $4740\,{\rm AU}$ and $4940\,{\rm AU}$, respectively, 
with final median separations of only $700\,{\rm AU}$ and $556\,{\rm AU}$, 
respectively. The medians are much smaller
than even the projected median separations in nearby clusters of 
$14800\,{\rm AU}$
\citep{gutermuth09}. The median separations in both simulations increased
by a factor of 3 or more between $t/\tff = 2.2$ and 2.4, so further evolution
may lead to larger mean separations.
In addition, the overall distributions, especially for G10, 
do resemble the observations (e.g. Fig. 2 of \citealt{gutermuth09})
in having a tail toward large separations. Given that the confusion
limit of the observations is about $3000\,{\rm AU}$, they would not
resolve the very close pairs.
Indeed, the peak of the distribution below $100\,{\rm AU}$ might be interpreted 
as binaries and multiple systems. 

\subsection{Evolution of Luminosity and Star Formation Efficiency }

Extragalactic observers employ a relation between total far-infrared
luminosity and star formation rate to estimate star formation rates
in dusty starbursts \citep{Kennicutt1998}:
\begin{equation}\label{eq:K98}
\textrm{SFR} (\msun \textrm{ yr}^{-1}) = 4.5 \ee{-44} L_{\rm{FIR}}
(\textrm{erg s}^{-1}) = 1.7 \ee{-10} L_{\rm{FIR}} (\lsun).
\end{equation}
For convenience, we express this K98 relation
as ${\rm SFR}/L = 1.7\ee{-4}\msun\,{\rm Myr}^{-1}\lsun^{-1}$ or
$L/{\rm SFR} = 5.9\ee3\lsun\,{\rm Myr}/\msun$.
A more recent calibration using the IMF of \citet{kroupa} yields
a slightly slower SFR, or higher
$L/{\rm SFR} = 6.9\ee3\lsun\,{\rm Myr}/\msun$.

\citet{wu2005} have used this relation for massive dense clumps in our
galaxy, finding a similar relation between $L_{\rm FIR}$ and the line
luminosity
of dense gas tracers like HCN for individual clumps as was found in starburst
galaxies by \citet{Gao04}, as long as $L_{\rm FIR}$ is above about
$\eten{4.5}\lsun$.
However, \citet{KT07} noted that the most massive stars, and hence the
luminosity, take a considerable time to build up during cluster formation.
Using an analytical prescription for star formation, they plotted
$L/{\rm SFR}$ versus time, finding values 1-2 orders of magnitude below
the extragalactic relation for times less than $1\,{\rm Myr}$.

\begin{figure}
\includegraphics[width=6in]{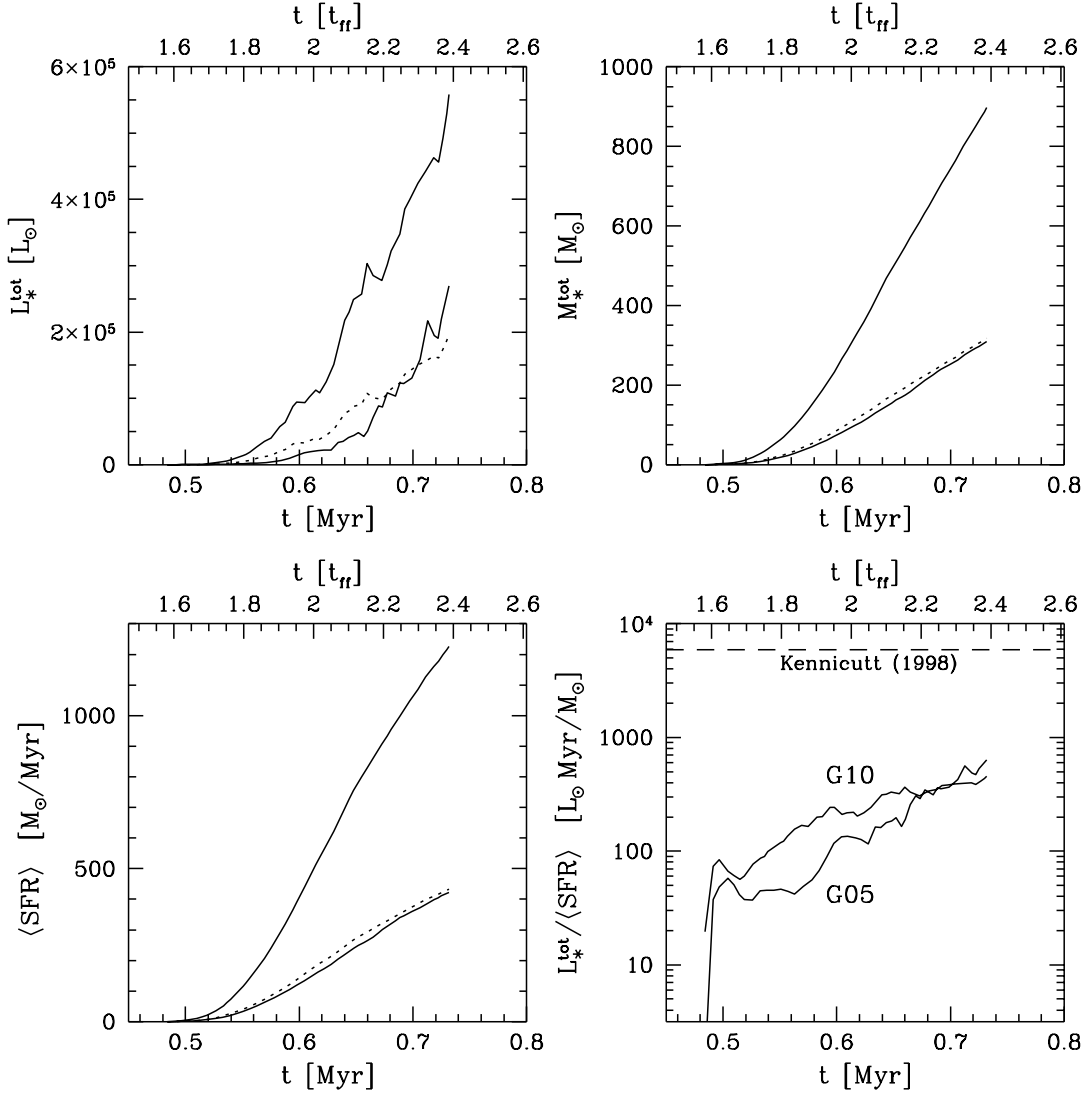}
\caption{Total luminosity (top left), total stellar mass (top right), 
average star formation rate (bottom left), 
and luminosity per average star formation rate (bottom right)
as a function of time for runs G05 and G10.
All panels show evolution of a different variable as a function of time, 
in both absolute years (bottom axes) and free-fall time (top axes). 
The top solid lines, bottom solid lines, and dotted lines correspond to 
run G10, run G05, and run G10 scaled
down by a factor of $2^{3/2}$, respectively.
Also plotted in the bottom right panel as a horizontal line labeled Kennicutt98
is the relation given in equation~(\ref{eq:K98}).}
\label{fig:sfr2}
\end{figure}

In Figure \ref{fig:sfr2}, we show how the total
luminosity $L_*^{\rm tot}$, total sink  mass $M_*^{\rm tot}$, 
time-averaged star formation rate
$\langle{\rm SFR}\rangle\equiv M_*^{\rm tot}/t$, 
and $L_*^{\rm tot}/\langle{\rm SFR}\rangle$ vary with time in our
simulations. 
The build-up of stellar mass and luminosity is quite rapid, but
the luminosity lags the mass.
The calculations were stopped at $2.5\tff$ or 0.75 Myr,
before the peak of star formation (see Fig. \ref{fig:sfr2}).
Both the sink formation rate and the mass accretion rate onto sinks are
still increasing at the end of the simulation (see Figure \ref{fig:gfrac1}).
At that time, $L_*^{\rm tot}/\langle{\rm SFR}\rangle$ 
lies a factor of 10 below the K98 relation.
When the total luminosity first exceeds $10^5\lsun$, the value
of $L_*^{\rm tot}/\langle{\rm SFR}\rangle$
lies about a factor of 20 below the K98 relation.
The values of $L_*^{\rm tot}$, $M_*^{\rm tot}$, and $\langle{\rm SFR}\rangle$
are larger for run G10 simply because the volume simulated is
larger. To allow a direct comparison with run G05, we plot the results for
run G10, scaled down by a factor of $2^{3/2}$ (dotted line). The results are
very similar to the ones for run G05.

These results confirm the prediction by \citet{KT07} that the K98
relation will underestimate the SFR for an individual cluster at early times,
but the discrepancies are somewhat less than they found. If star formation has
proceeded for a few free-fall times, or about $1\,{\rm Myr}$, 
the K98 relation
becomes better for an individual clump. However, there must still be
sufficient dust to convert most of the luminosity into far-infrared
radiation. Once ionizing radiation turns on, $\lfir$ may become 
a better tracer of the SFR. \citet{ve12} 
found agreement to a factor of 2 between SFRs calculated from $\lfir$ 
and radio continuum emission for a sample of massive dense clumps.

The last column of Table \ref{tbl:sim}
shows the star formation efficiency per free-fall time,
$\sfrff$, defined in \citet{KM05}. It is essentially the ratio
$f_{\rm sinks,final}/\tff$. The values are similar for all runs,
even though the number of sink particles is higher in
the new simulations compared to the simulation with only dust heating 
energetics. The values in Table \ref{tbl:sim} assume that all mass
entering the sink winds up in the star. As noted earlier, this almost
certainly overestimates the stellar mass by factors of 2-3, suggesting
values of $\sfrff$ of 0.07 to 0.10. Star formation efficiencies in massive
dense clumps are not well-constrained, but are certainly much lower.
Using the numbers for the mean $\lfir/M_{\rm vir}$ for massive dense clumps
from \citet{wu2010} and equation~(\ref{eq:K98}), along with the free-fall
time at the mean density of these clumps (0.27 Myr), yields values
of $\sfrff$ around 0.006. If $\lfir$ underestimates the SFR in those
clumps by a factor of 20, the values would agree better with those in 
Table \ref{tbl:sim}, but more likely the $\sfrff$ is lower than found
in our simulation.
Overly fast star formation is a common feature of simulations that do not 
include means to slow down the star formation process.
Even recent simulations with radiative feedback, turbulence, and outflows 
\citep{krumholz12}
produce very similar values of \sfrff\ to those in Table \ref{tbl:sim}.

\subsection{Density Profiles around Sinks}\label{sec:ddpe}

\begin{figure}
\includegraphics[width=6in]{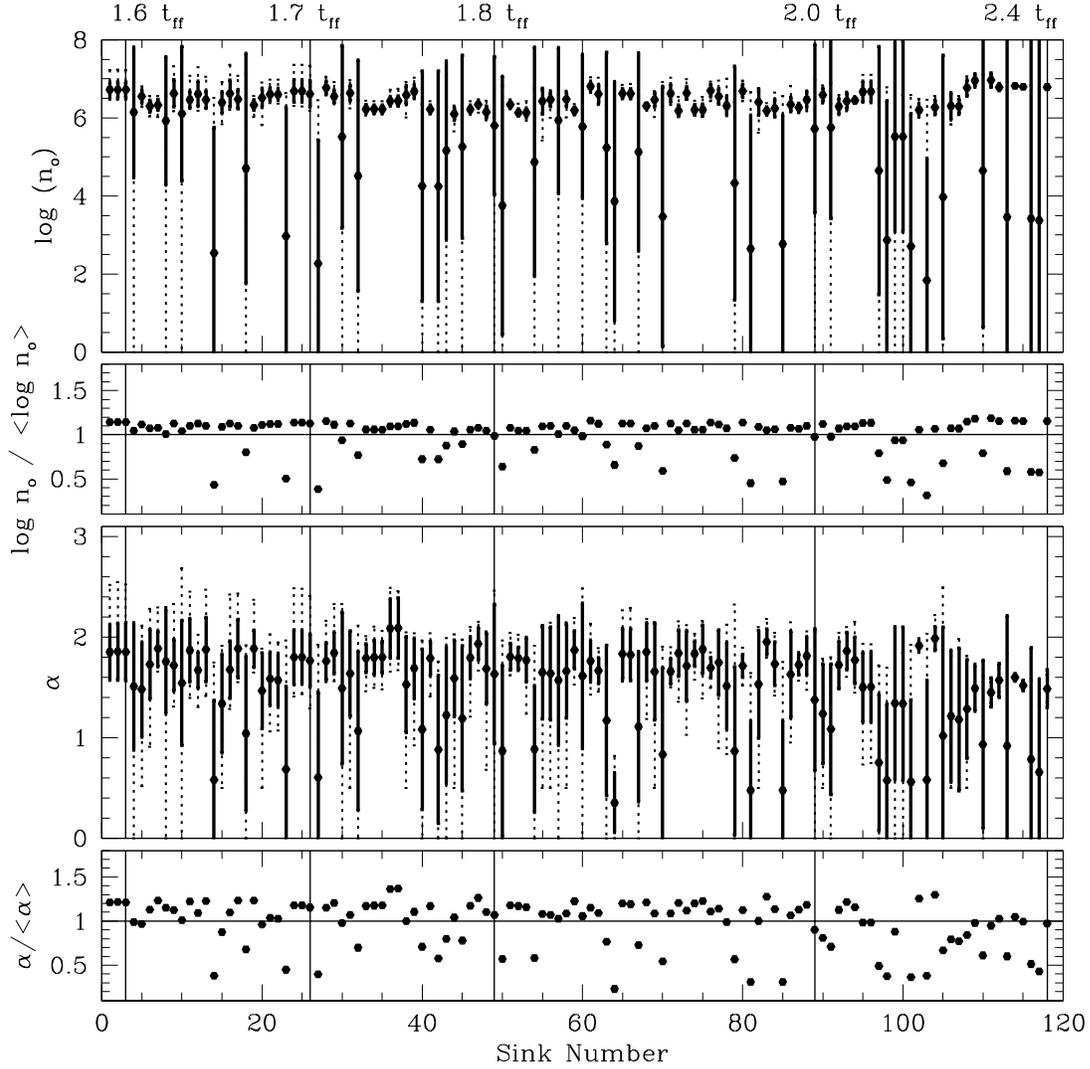}
\caption{Density profile parameters for all 118 sinks formed in simulation G05.
Values of density profile parameters, $\alpha$ and $n_o$, 
are shown for the sinks.
Error bars shown with a solid line indicate 
the standard deviation for each individual sink.
Error bars shown with a dotted line indicate the 
minimum and maximum value that $\alpha$ and $n_o$ take during the course
of the simulation.}
\label{fig:gdpe2d}
\end{figure}

We calculate the density profile around each sink at each time step.
The density profile will change around a sink as it moves within the cloud,
gathering gas particles and/or as gas particles accrete irregularly onto it.
In \S\ref{sec:code} we defined the density parameters, $n_o$ and $\alpha$, of
the density profile, equation (\ref{eq:n}).
Figure \ref{fig:gdpe2d} shows the average values of
the density profile parameters, $\alpha$ and $n_o$, for all the sinks that
formed in run G05. We only included sinks that have accretion rates
$\dot M>10^{-8}\msun\,{\rm yr}^{-1}$, because sinks with lower accretion rates
have too few neighboring gas particles to allow for an accurate
determination of the profile.
We summarize the results in Table \ref{tbl:dp}, together
with the results for run D05, taken from UME10.
The values are similar for both runs.  
This suggests that the density profile surrounding the individual
sinks is not strongly affected by the energetics algorithm used.

\begin{deluxetable}{lcccc}
\tablecaption{Density Profile around Sinks}
\tablewidth{0pt}
\tablehead{\colhead{Run} & \colhead{$\langle\alpha\rangle$} & 
\colhead{$\langle\log(n_o/{\rm cm}^{-3})\rangle$} }
\startdata
D05 & $1.7\pm0.4$ & $6.5\pm0.3$ \cr
G05 & $1.7\pm0.4$ & $6.4\pm0.3$ \cr
\enddata
\label{tbl:dp}
\end{deluxetable}

Groups studying young star-forming cores have calculated density profiles from
observations.
Comparing our average density profile values to those observationally derived
values, we find excellent agreement with our values of $\alpha$ and $n_o$.
\citet{shirley} studied Class 0 cores and found $\langle{\alpha}\rangle=1.63
\pm 0.33$ and a typical value of $\alpha=1.8\pm 0.1$ if they ignored two
sources with aspherical emission contours.
\citet{young} studied Class I cores and found $\langle{\alpha}\rangle=1.6 \pm
0.4$.
Our simulations are consistent with either of these values for $\alpha$.
The values of log $n_o$ derived from these two studies are $\langle
\textrm{log} (n_o/{\rm cm}^{-3})\rangle =6.1\pm0.2$ \citep{shirley} and
$\langle \textrm{log} (n_o/{\rm cm}^{-3})\rangle=5.4\pm0.5$ \citep{young}.
Our value of $\langle\textrm{log}(n_o/{\rm cm}^{-3})\rangle$ is
$6.4\pm0.3$, including only
points with accretion rates greater than $10^{-8}\msun\,{\rm yr}^{-1}$.  
These values
tend to agree better with the results for the Class~0 core study of
\citet{shirley}, suggesting that most of the sinks are reflecting early
stages in star formation.

\subsection{Evolution of Temperature, Density and Velocity}\label{sec:tempdens}

\begin{figure}
\includegraphics[width=6in]{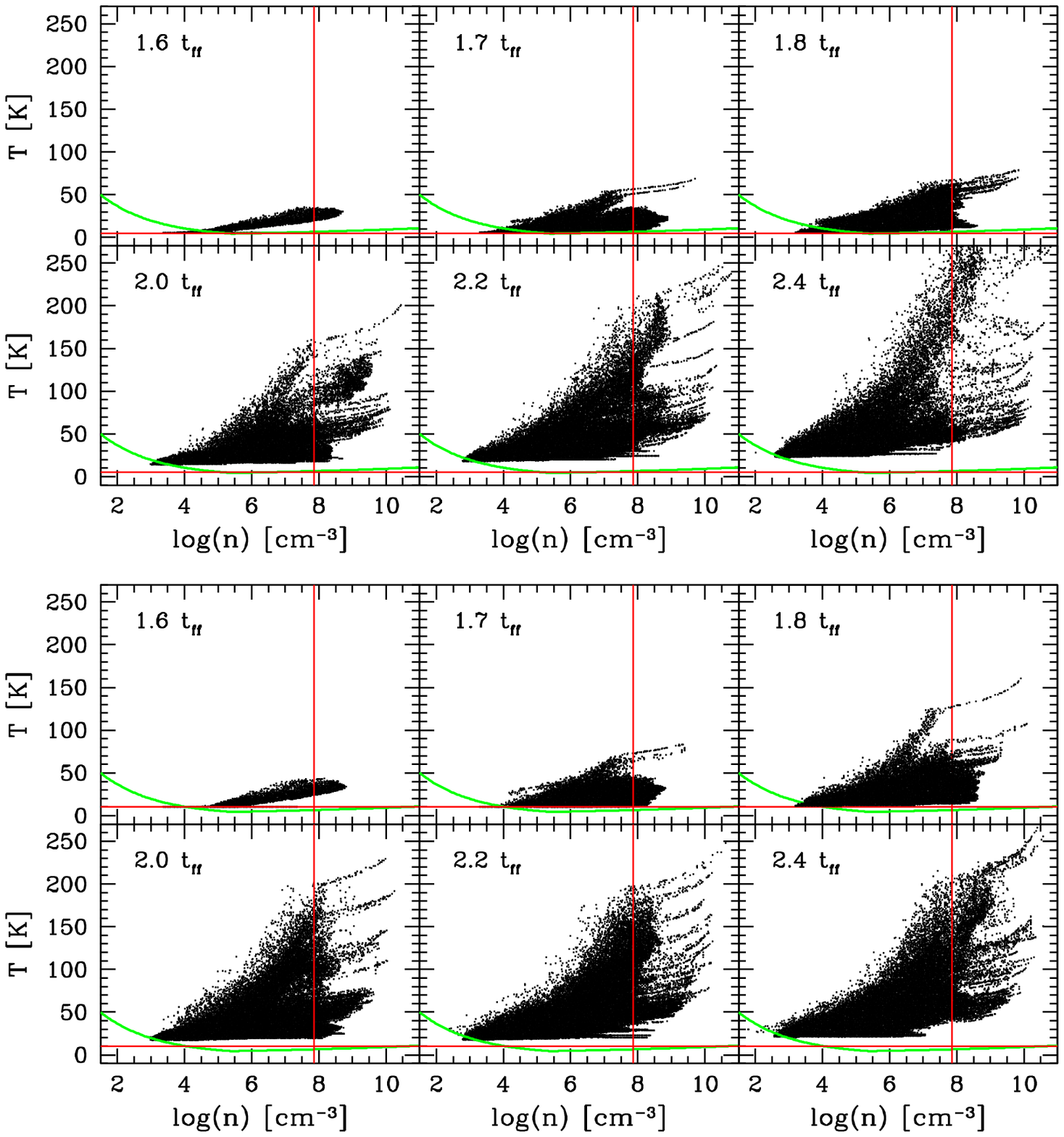}
\caption{Temperature and density of gas particles as a function of 
time for runs G05 (top 6 panels) and G10 (bottom 6 panels), at various
times, as indicated.
The green line shows the equation of state given by \citet{larson}.
The horizontal and vertical red lines show the minimum temperature $\tmin$
and threshold density $n_c$ for sink formation, respectively.
The statistics of the temperature are listed in Table~\ref{tbl:tempstats1}.}
\label{fig:gtd2}
\end{figure}

Figure~\ref{fig:gtd2} shows the density and temperature
evolution of gas particles in our simulations. The behavior of the gas
particles in these figures resembles what was seen in similar figures in UME10.
For example, at early times the temperatures of the gas particles are confined
to low values at all densities. However, as the simulation evolves and more
sink particles form, the gas particles are heated and tendrils of gas particles
appear in Figure \ref{fig:gtd2}. They are reaching toward
high-density and high-temperature regimes. These tendrils show the behavior of
the gas density and temperature near luminous sink particles.
Another interesting feature seen in this figure is the slight increase of the
floor of the temperature distribution as a function of time.
This feature of an increasing temperature floor was also seen in run
D05 (UME10). In Figure~\ref{fig:gtd2}, we also plot the equation of
state given by \citet{larson}. 
As in run D05 (UME10), the gas does not follow this
equation. This is not surprising because the equation of state of
\citet{larson} does not take into account stellar heating. This has been
confirmed to be an important effect in \citet{krumholz}, \citet{offner}, and
UME10. 

The horizontal and vertical red lines in Figure~\ref{fig:gtd2} show the
minimum temperature $\tmin$ and the density threshold $\rho_c$, respectively.
Gas particles located on the right of the vertical line, in regions where sinks
form, all have temperatures significantly larger than $\tmin$. These particles
do not turn into sinks because their high temperatures make them
unbound [see eq.~(13) in MES06].
This explains the results that were presented in Figure~\ref{fig:gfrac1} and
Table~\ref{tbl:sim}.
Increasing the minimum temperature from $5\,{\rm K}$ to $10\,{\rm K}$
might affect the formation
of the first few sinks, but once these sinks have formed 
and started reheating the
gas, the formation of subsequent sinks is unaffected by the particular value
of $\tmin$, because the gas temperature in regions of sink formation is 
already much larger. This explains why the gas fraction drops at essentially
the same rate in simulations G05 and G10 (Fig.~\ref{fig:gfrac1}), and why
both simulations form roughly the same number of sinks {\it per unit volume}\/.
Judging from Figure~\ref{fig:gtd2}, it would take a minimum
temperature of order $50\,{\rm K}$ 
or more to make a difference. 

Most of the cores in our simulations are located inside dense filaments.
The few cores located in low-density regions actually
formed inside filaments, and
were later ejected by 3-body encounters (MES06). This concentration of cores
can explain the large gas temperatures found in these
regions (even though more cores means more competition for accreting gas
from the same region, possibly leading to lower accretion luminosities).
The spatial distribution of cores is a consequence of the assumed
periodic boundary conditions.
\citet{heitschetal08} performed simulations of cloud fragmentation
with isolated boundary conditions, and showed that these initial conditions
lead to stronger initial fragmentation in the early stages,
and a more distributed core formation at late stages.
Observations are, however, finding strong concentrations of cores and
protostars along narrow filaments (e.g., \citealt{2012A&A...543L...3H}), 
very similar to patterns seen in the simulations.

\begin{deluxetable}{cccccrlc}
\tablecaption{Particle Statistics for Runs with Heating}
\tablewidth{0pt}
\tablehead{\colhead{Run} & \colhead{time $[\tff]$} & 
\colhead{$\tg\,[{\rm K}]$} & \colhead{$\td\,[{\rm K}]$} &
\colhead{$\log\,n\,[{\rm cm}^{-3}]$} & \colhead{$N_{\rm sinks}$} & 
\colhead{$f_{\rm sinks}$} & \colhead{$\max(M_{\rm sink})$} }
\startdata
D05 & 1.6 & $5.40\pm1.91$ & $5.40\pm1.91$ & $4.65\pm0.67$ &  3 & 0.02\% &  0.07 \cr
D05 & 1.7 & $9.28\pm3.84$ & $9.28\pm3.84$ & $5.02\pm1.08$ & 16 & 0.4\%  &  0.72 \cr
D05 & 1.8 & $14.8\pm6.08$ & $14.8\pm6.08$ & $5.12\pm1.12$ & 28 & 2.4\%  &  2.11 \cr
D05 & 2.0 & $24.5\pm9.66$ & $24.5\pm9.66$ & $5.05\pm1.15$ & 53 & 12\%   &  7.99 \cr
D05 & 2.2 & $33.0\pm14.3$ & $33.0\pm14.3$ & $4.88\pm1.16$ & 70 & 28\%   & 12.40 \cr
D05 & 2.4 & $48.3\pm29.7$ & $48.3\pm29.7$ & $4.76\pm1.21$ & 71 & 45\%   & 17.99 \cr
\hline
G05 & 1.6 & $5.64\pm1.84$ & $4.50\pm2.20$ & $4.63\pm0.68$ &   3 & 0.03\% & 0.06 \cr
G05 & 1.7 & $7.61\pm4.14$ & $9.64\pm3.89$ & $5.16\pm1.15$ &  26 & 0.5\%  & 0.78 \cr
G05 & 1.8 & $10.8\pm7.22$ & $14.9\pm5.94$ & $5.41\pm1.20$ &  49 & 2.8\%  & 1.7  \cr
G05 & 2.0 & $20.6\pm10.8$ & $25.9\pm10.6$ & $5.40\pm1.19$ &  89 & 14\%   & 7.9  \cr
G05 & 2.2 & $28.2\pm18.1$ & $33.6\pm17.8$ & $5.19\pm1.21$ & 108 & 30\%   & 12.8 \cr
G05 & 2.4 & $35.6\pm25.6$ & $43.0\pm25.4$ & $4.89\pm1.18$ & 118 & 46\%   & 20.8 \cr
\hline
G10 & 1.6 & $10.2\pm1.67$ & $5.35\pm2.60$ & $4.65\pm0.67$ &   5 & 0.03\% & 0.12 \cr
G10 & 1.7 & $11.4\pm4.24$ & $11.8\pm4.78$ & $5.21\pm1.11$ &  37 & 0.6\%  & 2.22 \cr
G10 & 1.8 & $17.5\pm7.07$ & $19.6\pm7.11$ & $5.63\pm1.26$ & 138 & 3.1\%  & 6.08 \cr
G10 & 2.0 & $26.7\pm12.8$ & $30.0\pm13.1$ & $5.76\pm1.22$ & 301 & 15\%   & 15.1 \cr
G10 & 2.2 & $33.1\pm18.2$ & $35.7\pm18.0$ & $5.57\pm1.22$ & 346 & 32\%   & 21.8 \cr
G10 & 2.4 & $39.6\pm22.6$ & $42.7\pm22.5$ & $5.41\pm1.24$ & 365 & 47\%   & 24.0 \cr
\enddata
\label{tbl:tempstats1}
\end{deluxetable}

Table \ref{tbl:tempstats1} shows various statistics for runs with heating 
(D05, G05, and G10), i.e., the average gas and dust temperatures, 
and average density (calculated by averaging over all gas particles),
number of sink particles, percentage of mass in sink particles, and 
mass of the most massive sink particle, at different times. We find that
the average gas and dust temperatures both increase throughout the simulations.
This increase can be explained by the increasing 
number of sink particles and the increasing mass of the individual sink 
particles, both of which lead to more substantial radiation fields.

\begin{figure}
\epsscale {1.0}
\includegraphics[width=6in]{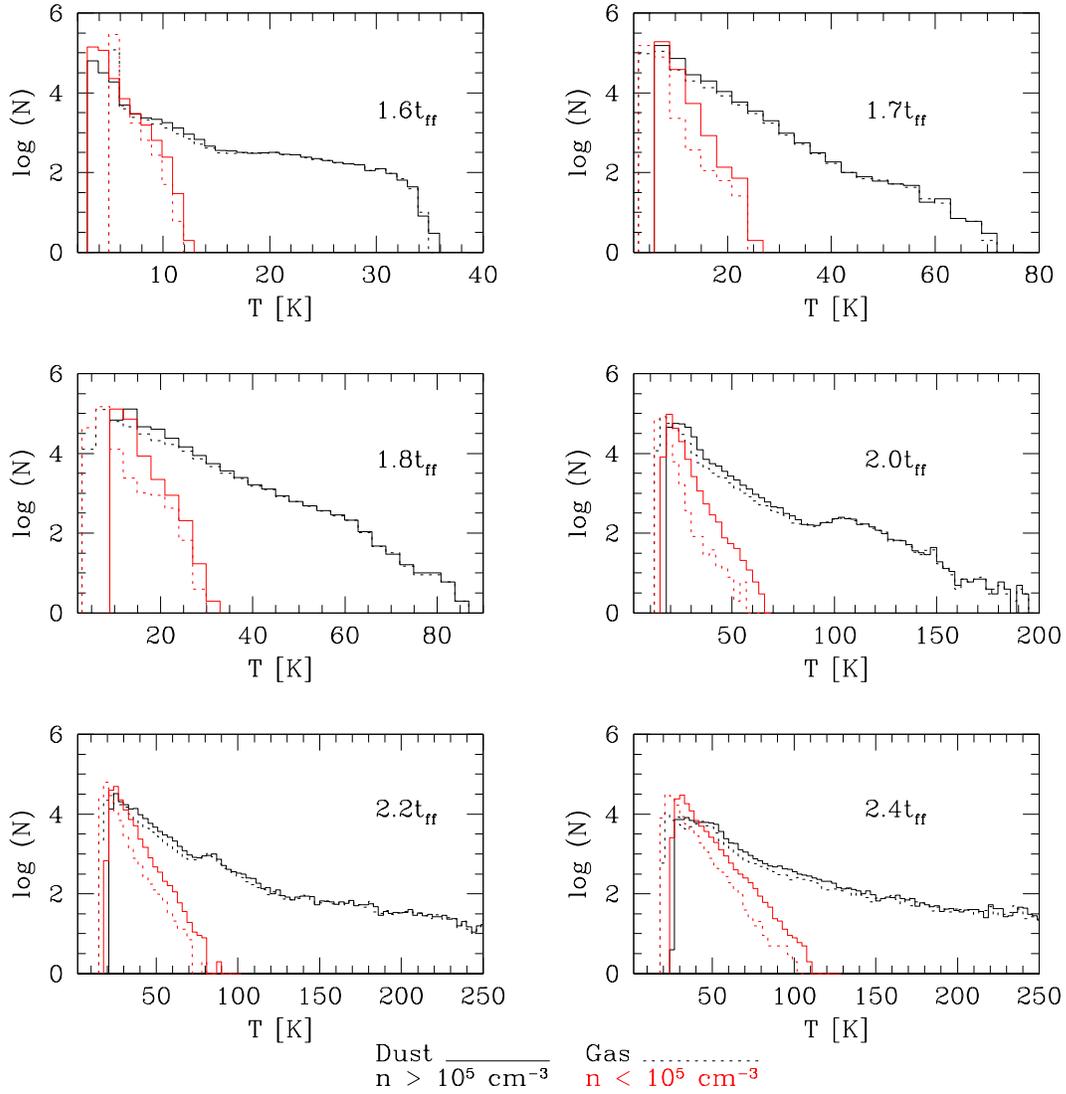}
\caption{Dust and gas temperature histograms in the low and high 
density regimes, for run G05.  
The dust (solid line) and gas (dotted line) temperature histograms are
plotted for low (red) and high (black) densities.}
\label{fig:t2hist}
\end{figure}

Since it is essentially the mass of sinks and accretion rate of gas onto
sinks that determine the dust temperature, we expect the temperature of the
gas being accreted to have
little effect if it accretes supersonically, which is indeed the case (see
Table~\ref{tbl:lmae} below).
At all times $t\geq1.7\tff$ in both G05 and G10, the mean
gas temperature is slightly cooler than the dust temperature.  
This is expected since some gas is at quite low densities ($n \sim
\eten3$ or \eten4 \cmv) and the rest is no hotter than the dust
temperature.
The gas is fully equilibrated to the dust temperature only at 
quite high densities, such as $n > \eten6$ \cmv.
The dust temperature is initially fixed at $5\,{\rm K}$ for run G05
and $10\,{\rm K}$ for run G10, 
and it can change only when
heating via nearby stellar radiation fields begins.
Figure \ref{fig:t2hist} illustrates these effects in more detail. It shows the
histograms of the dust and gas temperature at various times for low-density
($n < \eten5\,\cmv$)  and high-density ($n > \eten{5}\,\cmv$) gas.  
Except for the first time step shown, the dust
temperature is higher than the gas temperature at low densities (red
histograms). It also shows that the dust
and gas temperatures are nearly equal at high densities (black histograms)  due
to dust-gas collisional coupling. At $t=1.6\tff$,
the dust is cooler on average than the gas, for low density
gas (see also, the first line of Table~\ref{tbl:tempstats1}).
At these early
times, the dust has not yet been heated to high temperatures by forming stars.
Because the dust temperatures are lower, the dominant heating source for
low-density gas in our simulations is cosmic-rays, 
which can raise the gas temperature above
the dust temperature.

\begin{figure}
\includegraphics[width=6in]{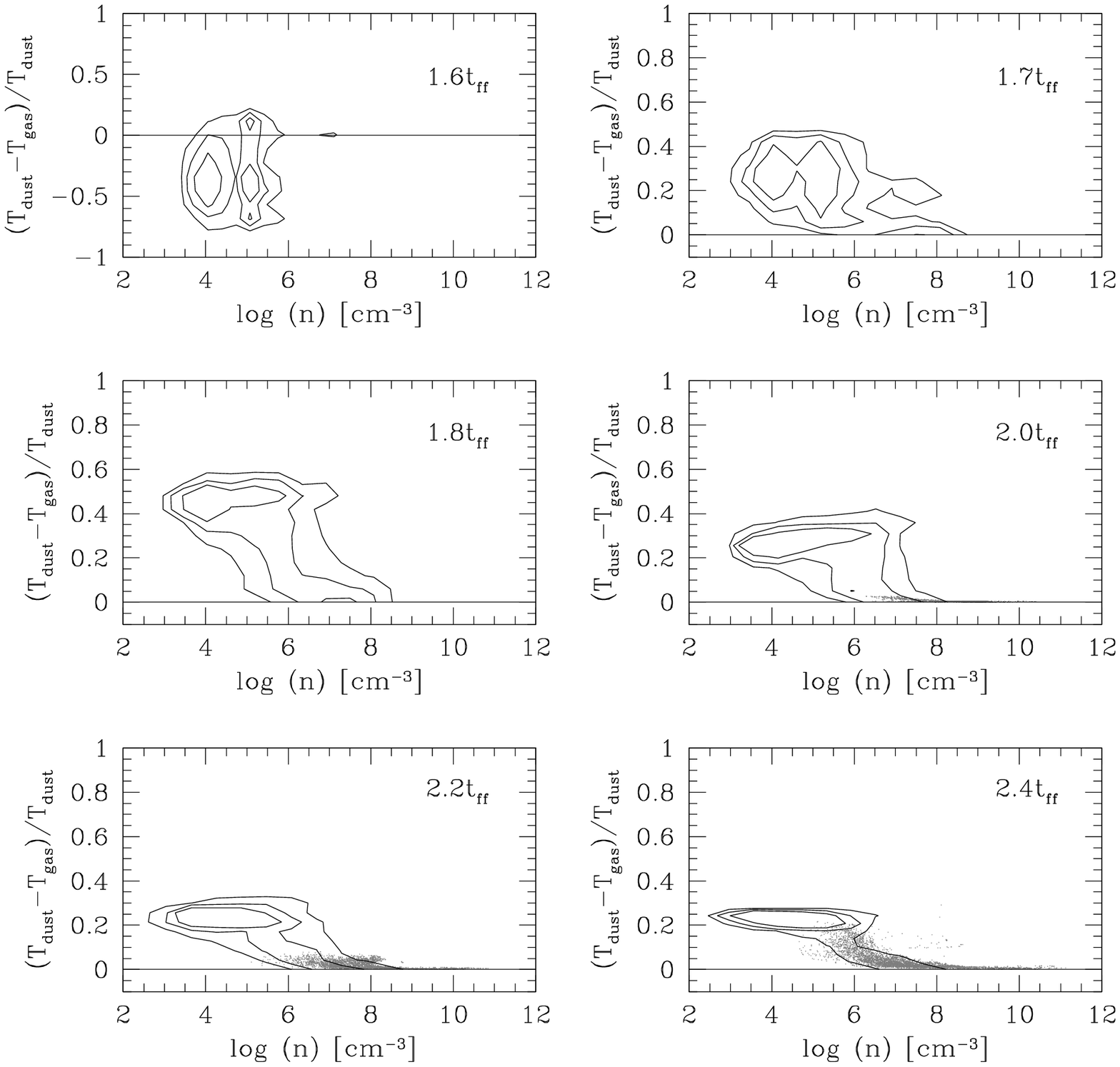}
\caption{Temperature difference versus density shown at various times for 
run G05. Horizontal line marks equal dust and gas temperature.
Contours show 3, 10, and 20 $\sigma$ contour levels of density of SPH
gas particles. Dots indicate gas particles which have dust temperatures
greater than $100\,{\rm K}$.  
}
\label{fig:gtgtd2}
\end{figure}

Figure \ref{fig:gtgtd2} shows the dust-gas temperature
difference as a function of density at various times for run G05.
The contours indicate the region in parameter space,
$\td-\tg$ versus $\log\,n$, where the majority of the
material is located. The gas is confined to the
same region for most of the simulation, except during early times when the dust
is heating up due to formation of luminosity sources. As seen before in Table
\ref{tbl:tempstats1} and Figure \ref{fig:t2hist} 
during the earliest times, the gas temperature is hotter than
the dust temperature because of the low-intensity of the radiation field.
At high densities ($\log\,n>8$), the SPH gas particles tend to approach the
horizontal line which indicates equal dust and gas temperature. At these
high densities when the dust and gas temperatures are comparable, dust heating
dominates all other forms of heating/cooling for the gas. This result was seen
in UED09. By $2\tff$, almost all the gas has $\tkin$ within 30\% of
$\td$, and a large fraction is even closer to equilibration.
Figure \ref{fig:gtgtd2} also shows
that high dust temperatures are most likely to exist in high-density regions
(as seen by the points), presumably due to the proximity of a nearby
forming star.

Figures \ref{fig:t2hist} and \ref{fig:gtgtd2} show that neither the
gas nor the dust are well described by isothermal or barotropic equations
of state once protostars begin to heat their surroundings. There is
a strong effect on the thermal behavior of the clump once protostars
develop significant luminosity. That effect occurs during the sink
formation process.

\begin{deluxetable}{lccc}
\tablecaption{Statistics of Accretion onto Sinks.}
\tablewidth{0pt}
\tablehead{\colhead{Run} & \colhead{${\cal M}$} &
\colhead{$v\,[{\rm km\,s^{-1}}]$}
& \colhead{$\dot M\,[10^{-5}\msun\,{\rm yr}^{-1}]$} }
\startdata
D05 & $10.8\pm7.4$ & $9.2\pm8.2$ & $2.20\pm3.08$ \cr
G05 &  $8.2\pm5.9$ & $5.5\pm5.3$ & $1.65\pm2.32$ \cr
G10 & $10.8\pm7.6$ & $6.5\pm6.0$ & $1.37\pm2.36$ \cr
\enddata
\label{tbl:lmae}
\end{deluxetable}

For each gas particle accreted onto a sink, we calculated the velocity
$v$ of the particle relative to the sink, and the Mach number ${\cal M}=v/c_s$,
where $c_s$ is the sound speed at the location of the sink.
Table~\ref{tbl:lmae} shows the mean values of $v$ and $\cal M$, and also the
mean values of the accretion rate
$\dot M$, for all simulations with heating. Again, similar
behavior of the sinks is seen for both G05 and G10, and also D05. 
On average, particles are being accreted supersonically.
The sound speeds near sinks are comparable for runs G05 and G10, since,
as we saw, the temperature in these regions greatly exceeds $\tmin$.
Hence, the slightly larger accretion velocities for run G10 results
in slightly larger Mach numbers. Compared to the runs with full
energetics, run D05 has on average larger accretion rates, larger accretion
velocities, and comparable Mach numbers. This run forms fewer sinks, but
the total mass in sinks is the same (see Fig.~\ref{fig:gfrac1}
and Table~\ref{tbl:sim}). Hence, the
same mass is being accreted by fewer sinks, resulting in higher accretion rates.
Also, fewer sinks means that they are on average more massive. Since the
accretion velocities are highly supersonic, gas particles accrete essentially
at the escape velocity, which is larger for more massive sinks. 
However, the gas temperature, and
consequently the sound speed, is also larger for run D05. As a result,
the Mach numbers are comparable to the ones for runs G05 and G10.

These infall velocities are considerably higher than the assumed
turbulent broadening (Doppler $b$ parameter is set to $1\,\kms$ 
as discussed in \S\ref{sec:tdust}). 
To the extent that these larger velocities allow
for greater escape probabilities for photons in the primary cooling
lines, the gas could be somewhat cooler than we calculate in the
infalling gas around a sink.
On the other hand, these are just the regions where the density becomes
large enough to overwhelm gas cooling and couple the gas temperature to
the dust temperature, so the effects are probably minimal.

\subsection{Mass Functions}\label{sec:MF}

The Galactic Field Star IMF has been studied by many groups (see
\citealt{salpeter}; \citealt{ms79}; \citealt{scalo}; \citealt{kroupa};
\citealt{chabrier}).
The cluster IMF has also been studied in young clustered,
star-forming regions such as Orion \citep{hill97} and in 
more isolated star-forming regions in the Taurus
\citep{briceno02}, Lupus \citep{Comeron2009}, and Chamaeleon \citep{Alcala1997}
molecular clouds.    In a recent review paper, \citet{Bastian2010} found
no significant variations in the IMF for present-day star formation.  

In large-scale simulations of cluster formation, various
treatments of thermal energetics were used. The earliest attempts assumed an
isothermal equation of state (\citealt{klessen}; MES06).
In the work of MES06, we found that a properly-resolved isothermal
calculation could only produce very low-mass fragments. The mass function in
these simulations was log-normal with a peak that was determined by the
resolution limit of the simulation. Other simulations attempted to address
this issue by modifying the equation of state to account for the increase in
optical depth in higher density gas. They used a barotropic
equation of state in which the temperature of the gas increases at higher
densities \citep{bbb,li,bate05,jappsen,larson,bcb,clark}. In more
recent work, groups have included various treatments 
of radiative transfer to more
accurately account for the heating effect of the young stars on the gas 
(\citealt{krumholz10,bate09r,offner,smith}; UME10).
Inclusion of the heating effect shifts the mass function to higher masses.

\begin{figure}
\includegraphics[width=6in]{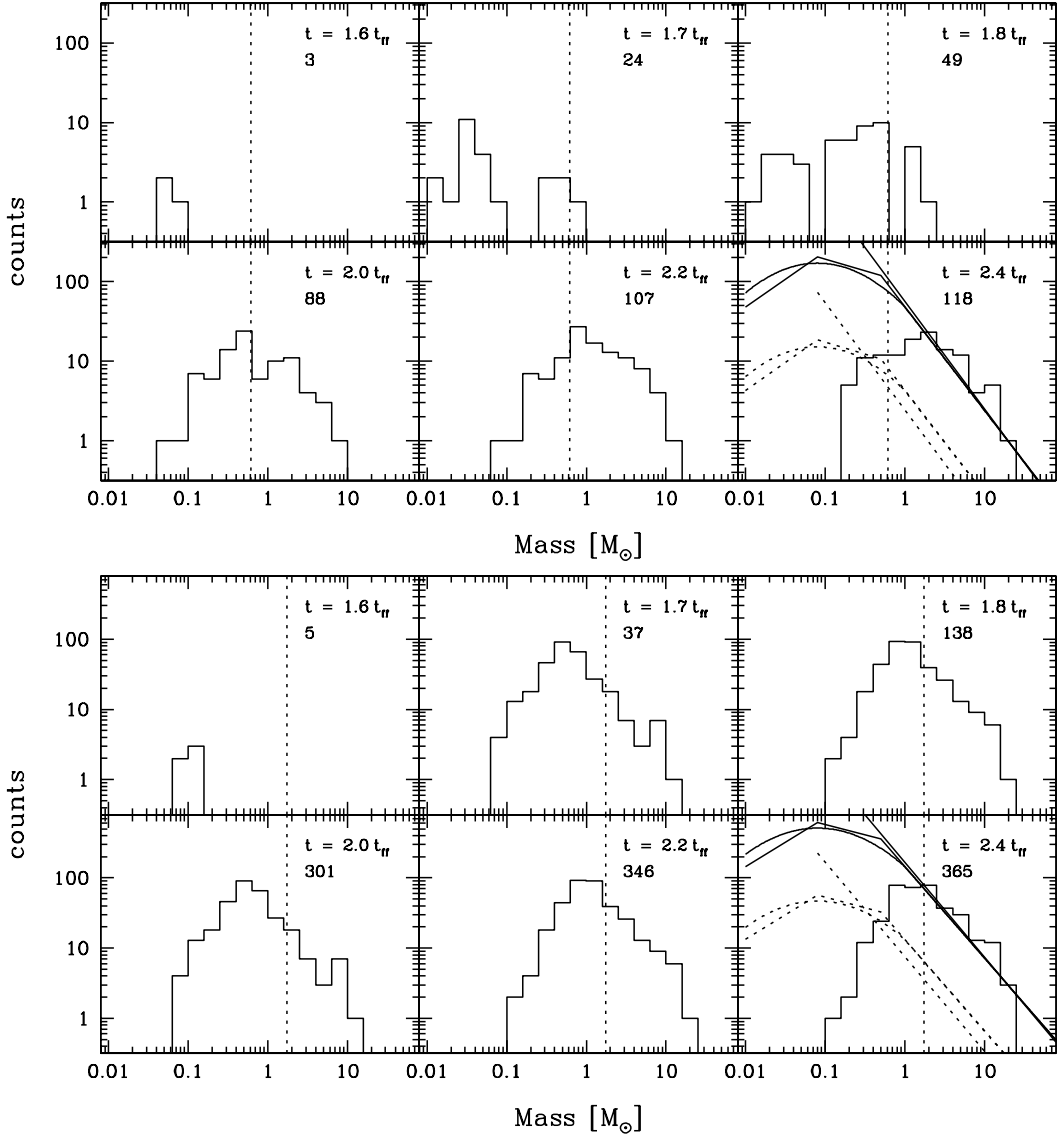}
\caption{Mass histograms for runs G05 (top panels) and G10 (bottom panels), 
shown at different times. The time and number of sinks are indicated in
each panel. The analytic mass functions of
\citet{salpeter} (straight line), \citet{chabrier} (curved line), 
and \citet{kroupa} 
(segmented line) are shown 
as solid lines when normalized to maximum mass object,
and dashed lines when
normalized to total number of objects. 
Vertical dashed lines show initial Jeans mass in the simulation.
}
\label{fig:gmf2}
\end{figure}

Figure \ref{fig:gmf2} shows the mass functions of the sink
particles at various times in the simulations G05 and G10. 
At early times, the mass function is dominated
by accreting low-mass objects. As the simulations evolve, these low-mass
sinks continue to accrete and become more massive. At later times, the
formation of low-mass sinks become less frequent because
the gas is now hotter and the Jeans mass is higher,
causing  a delay in sink formation until they have reached the new
higher Jeans mass. Although there is a tendency for the mass function
to shift its peak to higher masses with time, the IMF in simulation G10
becomes relatively stable after about $1.8\,\tff$.
The shift to higher masses relative to the isothermal calculation
agrees with the results of the references given above.

We also plot the IMF's from
\citet{salpeter}, \citet{chabrier}, and \citet{kroupa} in the last panel
for each run.
We normalize the mass functions using the maximum mass in the
simulation (solid curves) and the total number of objects (dashed curves), both
given in Table \ref{tbl:sim}.  The minimum mass used to create the IMF's was 
$0.08 \msun$ and $0.01 \msun$ for the \citet{salpeter} and 
\citet{chabrier}/\citet{kroupa} IMF's, respectively. Normalizing the mass
functions based on the total number of objects highlights the fact that we
form too few objects given the mass of the most massive sink in each
simulation. The IMF's normalized to the maximum mass 
demonstrate that we can roughly reproduce the slope of high-mass tail 
above masses of $\sim1\msun$.

\begin{figure}
\includegraphics[width=6in]{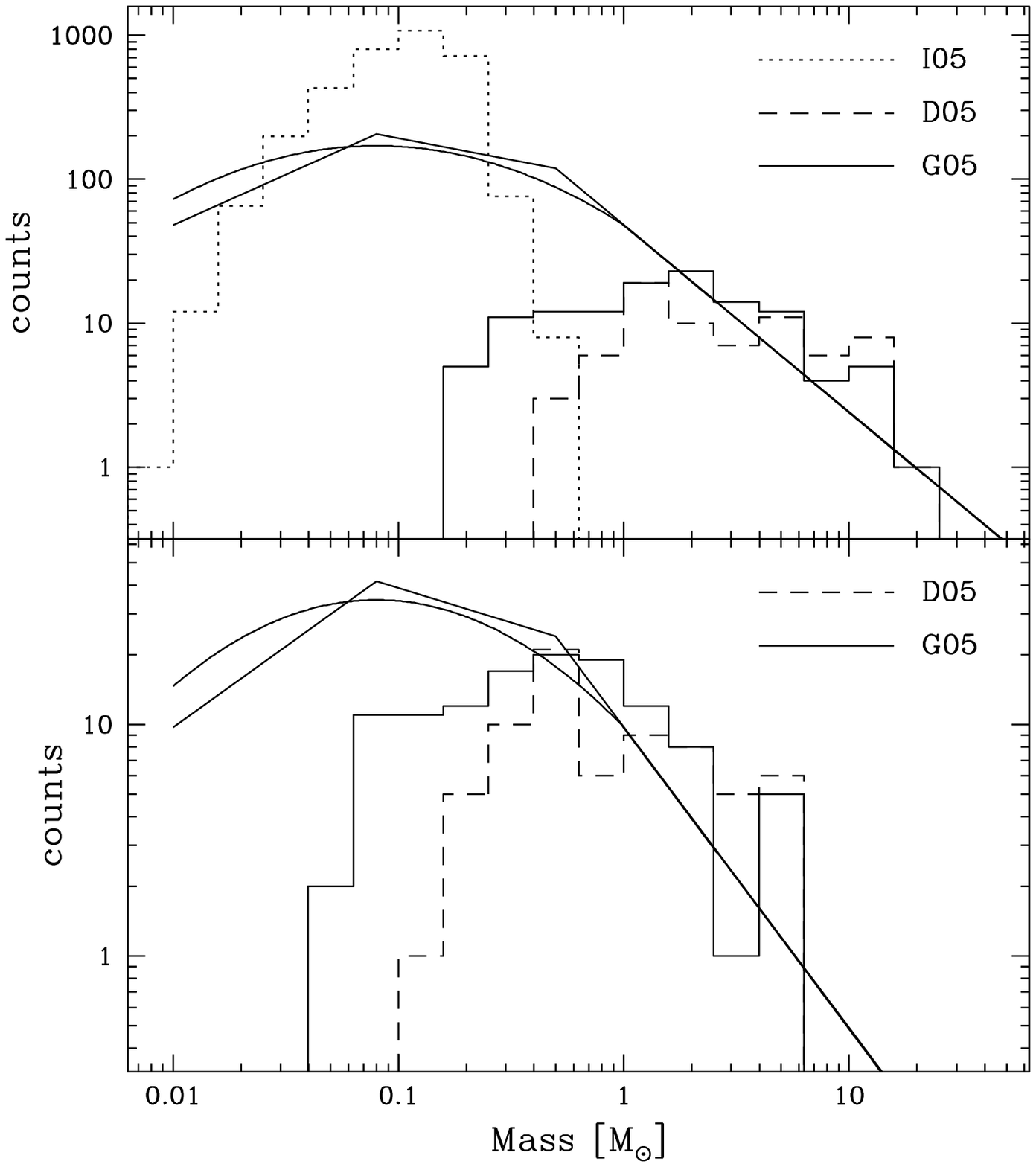}
\caption{Top panel:
mass histograms for all simulations with $\tmin=5{\rm K}$.  
Mass function are shown at final time ($2.5\tff$ for I05 and D05,
$2.4\tff$ for G05). 
The empirical IMF from \citet{chabrier} and \citet{kroupa} are shown,
normalized to a maximum mass of $\sim20\msun$.
Bottom panel: mass histograms for runs D05 and G05, 
and efficiency of 30\%.}
\label{fig:mfall}
\end{figure}

As in UME10, our mass functions slightly over-predict the number of high-mass
objects and miss a substantial fraction of low-mass objects.
In UME10, we predicted that by including molecular cooling and cosmic-ray
heating we would see a decrease in high-mass objects and an increase in
intermediate- to low-mass objects.
This has turned out to be the case and can be seen in the top panel
of Figure~\ref{fig:mfall}
where we show the mass functions for our three different simulations with
a minimum temperature of $5\,{\rm K}$.
We form more intermediate-mass objects, while still retaining
the slope at higher masses.  

\begin{deluxetable}{lcc}
\tablecaption{Average and Median Masses}
\tablewidth{0pt}
\tablehead{
\colhead{Run/observations} & \colhead{Median} & 
\colhead{Average}
}
\startdata
D05                              & 2.26 & 4.63 \cr
G05                              & 1.63 & 2.65 \cr
G10                              & 1.50 & 2.46 \cr
G10 ($\times 0.30$)		 & 0.45 & 0.74 \cr
\hline
\citet{salpeter}\footnotemark[1] & 0.13 & 0.26 \cr
\citet{chabrier}\footnotemark[1] & 0.10 & 0.31 \cr
\citet{kroupa}\footnotemark[1]   & 0.15 & 0.34 \cr
\enddata
\label{tbl:avgmed}
\tablenotetext{1}{Median calculated using normalization to total number 
of objects formed;
Average calculated using normalization to maximum mass
object.}
\end{deluxetable}

In Table \ref{tbl:avgmed}, we give the average and 
median sink masses in our simulations. We also give these values predicted by
various empirical IMF's. We use a lower mass limit of $0.08\msun$ for the
Salpeter IMF and $0.01\msun$ for the Kroupa and Chabrier IMF's.
Including the complete energetics algorithm (G05 and G10)
decreases the resulting mean mass, compared to D05, as predicted. 
The change in the median mass shows a similar behavior.
Although  our median and average masses decrease, they still give
values which are in disagreement with the empirical IMFs.
Next, we perform a thought experiment to address this issue.  

The regions around our sinks have density profiles similar to those
of gas around Class 0 objects discussed in \S \ref{sec:ddpe}. Thus, it
is reasonable to associate them with dense cores. As this material
falls into the sinks, we have so far assumed that all the mass winds
up in a star.
\citet{Alves2007} claim that the dense core mass
function is related to the stellar IMF, i.\ e., sharing the same shape, but
shifted to higher masses.  This is interpreted as a core-to-star efficiency
factor of 30\%. \citet{Enoch2008} find a lower-limit of
25\% for the core-to-star formation efficiency. The missing mass is likely
to be removed in molecular outflows driven by stellar winds (e.g.,
\citealt{Dunham2010, hansen12}). 
These outflows are not included in our simulation.
To include this inefficiency, we scale the mass of sinks in the data
from our simulations down by a factor of $0.30$.
We then recalculate the shape of the empirical stellar IMF for the new
maximum masses and total number of stars.  We show the results in the
bottom panel of Figure~\ref{fig:mfall}.   
We plot the empirical IMF normalized to the new maximum mass.  
Our results exhibit better agreement with a
stellar mass function with a 30\% efficiency factor.
Since we have only decreased the masses, the high-mass slope in our new 
simulation data remains the same and agrees with the empirical IMF.  
In addition, the peak in the IMF shows better agreement. 
At $0.1\msun$, our new mass function (for the complete energetics 
algorithm) is only missing a factor of 2-3 objects compared to the
empirical  IMF, 
which is significantly less than the factor of $\sim20$ seen in Figure
\ref{fig:mfall}. However, the adjusted mean and median masses
are still higher than in the empirical IMFs (Table \ref{tbl:avgmed}).
Since we assumed that
all the mass of the sink went into the star in order to calculate
luminosities, 
it is not self-consistent to apply the efficiency factor after the calculation.
We do it only to suggest avenues for future exploration.
There are questions about the connection between the core mass function
and the stellar mass function (e.g., \citealt{2007MNRAS.379...57C, smith}).
Other effects may decrease the mass of the final stars, including
disk fragmentation to form binary or multiple stars and brown dwarfs
(e.g., \citealt{2011MNRAS.413.1787S}).

\section{Summary and Conclusion}\label{sec:con}

The motivation for this work was to explore the effect of dust-gas energetics
in a clustered star formation simulation.  
We have presented the results of two new simulations (G05 and G10), which
include the complete energetics algorithm discussed in UED09.
The properties of the models, including two previous ones
(I05 and D05) are
summarized in Table \ref{tbl:params}, and we compare them here.

As in D05 and I05, sink particles, representing protostars,
form along filaments and especially at intersections of filaments.
The temperature of a simulation which included gas cooling was on average
lower than in a simulation with $\tk = \td$, as expected. With a lower
average temperature, fragmentation was more prevalent and more objects were
able to form.
The average density profile parameters surrounding a sink 
were similar among the four simulations and agreed with observations
of low-mass Class~0 sources. 
However the infall speeds were significantly supersonic,
and mass accretion rates were high, both
in contrast to observations of low-mass protostars. 
Infall speeds and mass accretion rates for
high-mass protostars and protostars in clustered environments are poorly
constrained. Infall speeds and mass accretion rates are somewhat smaller
for G05 and G10 compared to D05.

We added a calculation of the ratio of far-infrared luminosity ($\lfir$)
over the SFR to test the use of $\lfir$ as a SFR tracer in very young
regions of clustered star formation. We found that $\lfir/{\rm SFR}$ increases
rapidly during the simulation, but that it is significantly lower 
(factor of 10) than the ratio used to measure extragalactic SFR at the 
end of the simulations (around $0.7\,{\rm Myr}$). Measurements of SFR for very
young clusters (ages $<1\,{\rm Myr}$) using $\lfir$
are very likely underestimated.

We computed the mass evolution of protostars during the simulation,
and we compared the mass function at the end of each simulation.
In our previous work, UME10, we
found that a non-isothermal, stellar-source dependent energetics algorithm
radically reduced the number of young stars that were formed and formed
more massive stars, compared to simulations with isothermal gas.
However, the simulations in UME10 over-produced high-mass objects and missed a
large fraction of low-mass objects.  We predicted in UME10 that including a
more realistic calculation of the gas temperature might address this problem.
In this work, we included the complete dust-gas energetics algorithm.
This change increased the number of intermediate mass objects, but
the deficiency of low-mass objects persists.

The two main differences between D05 versus G05 and G10
were the  temperature distribution and the mass function,
which are related to each other.
In a lower temperature environment with more sink particles forming, there was
less material available to be accreted and therefore a smaller fraction of
massive objects were formed.   This affected the mass function and led to a
slight decrease in the number of high-mass objects and an increase in the
amount of low-mass objects when compared to the simulations with dust heating
only in UME10. We found very little difference in the mass function 
between G05 and G10, indicating that the initial temperature is not
very important; feedback from the first protostars rapidly erases the
effects of initial temperature.

We performed a thought experiment in which we tried to explain the discrepancy
between our mass function and the empirical IMF.  In our simulation, we 
assumed that all the mass in a sink particle winds up
in a single star. However, studies of nearby clouds (e.g., \citealt{Alves2007} 
and \citealt{Enoch2008}) show
that about 70\% of the core mass does not wind up in the star,
probably because it is removed by stellar winds and the resulting molecular
outflow (e.g., \citealt{Dunham2010}). If we multiply our sink masses by 0.3, we
get better agreement with the IMF.  
Although we have shown that including dust-gas energetics is essential,
other effects  (e.g., magnetic fields, turbulence, outflows, etc.) 
will need to be included for a full understanding.
In a promising development, \citet{hansen12} found that including outflows
along with radiative feedback
reduced the mass accretion rates and protostellar masses, hence luminosities,
allowing more fragmentation and better reproducing the IMF.
More recently, \citet{krumholz12} have found better agreement with the IMF
when turbulence, outflows, and radiative transport are included, although
their IMFs are still somewhat top-heavy.

\acknowledgments

All calculations were performed at the Laboratoire d'astrophysique
num\'erique, Universit\'e Laval.  We are pleased to acknowledge the support of
NASA Grants NAG5-10826, and NAG5-13271.  We would also like to thank the Canada
Research Chair program (H.M.), NSERC (H.M.), NSF Grants AST-0607793 and 
AST-1109116 (N.E.), and
the NASA GSRP Fellowship Program (A.U.) for providing support for this work.
Part of A. Urban's contribution to the research described in this publication
was carried out at the Jet Propulsion Laboratory, California Institute of
Technology, under a contract with the National Aeronautics and Space
Administration.


\begin{thebibliography}

\bibitem[Alcala et~al.(1997)]{Alcala1997}
Alcala, J.~M., Krautter, J., Covino, E., Neuhaeuser, R., Schmitt,
J.~H.~M.~M., \& Wichmann, R. 1997, \aap, 319, 184

\bibitem[Alves et~al.(2007)]{Alves2007}
Alves, J., Lombardi, M., \& Lada, C.~J. 2007, \aap, 462, L17

\bibitem[Bastian et~al.(2010)]{Bastian2010}
Bastian, N., Covey, K.~R., \& Meyer, M.~R. 2010, 
ARA\&A, 48, 339

\bibitem[Bate(2005)]{bate05}
Bate, M. R. 2005, MNRAS, 363, 363

\bibitem[Bate(2009)]{bate09r}
Bate, M.~R, 2009, \mnras, 392, 1363

\bibitem[Bate et~al.(2003)]{bbb}
Bate, M.~R., Bonnell, I.~A., \& Bromm, V. 2003, \mnras, 339, 577

\bibitem[Bonnell et~al.(2006)]{bcb}
Bonnell, I.~A., Clarke, C.~J., \& Bate, M.~R. 2006, \mnras, 368, 1296

\bibitem[Bressert et al.(2010)]{bressert10} 
Bressert, E. et al. 2010, \mnras, 409, L54 

\bibitem[Brice\~no et~al.(2002)]{briceno02}
Brice\~no, C., Luhman, K.~L., Hartmann, L., Stauffer, J.~R., \&
  Kirkpatrick, J.~D. 2002, \apj, 580, 317

\bibitem[Bromm et~al.(2002)]{bcl}
Bromm, V., Coppi, P.~S., \& Larson, R.~B. 2002, \apj, 564, 23

\bibitem[Chabrier(2003)]{chabrier}
Chabrier, G. 2003, \pasp, 115, 763

\bibitem[Clark et~al.(2008)]{clark}
Clark, P.~C., Glover, S.~C.~O., \& Klessen, R.~S. 2008, \apj, 672, 757

\bibitem[Clark et al.(2007)]{2007MNRAS.379...57C} Clark, P.~C., Klessen, 
R.~S., \& Bonnell, I.~A.\ 2007, \mnras, 379, 57

\bibitem[Comer\'on et~al.(2009)]{Comeron2009}
Comer\'on, F., Spezzi, L., \& L\'opez Mart\'{\i}, B. 2009, \aap, 500, 1045

\bibitem[Dale et~al.(2005)]{Dale2005}
Dale, J.~E., Bonnell, I.~A., Clarke, C.~J., \& Bate, M.~R. 2005,
  \mnras, 358, 291

\bibitem[Dunham et~al.(2010)]{Dunham2010}
Dunham, M.~M., Evans, N.~J., Terebey, S., Dullemond, C.~P., \& Young,
  C.~H. 2010, \apj, 710, 470

\bibitem[Enoch et~al.(2008)]{Enoch2008}
Enoch, M.~L., Evans, II, N.~J., Sargent, A.~I., Glenn, J.,
  Rosolowsky, E., \& Myers, P. 2008, \apj, 684, 1240

\bibitem[Gao \& Solomon(2004)]{Gao04}
Gao, Y., \& Solomon, P.~M. 2004, \apj, 606, 271

\bibitem[Goldsmith(2001)]{2001ApJ...557..736G} Goldsmith, P.~F.\ 2001, 
\apj, 557, 736 

\bibitem[Gutermuth et~al.(2009)]{gutermuth09}
Gutermuth, R.~A., Megeath, S.~T., Myers, P.~C., Allen, L. E.,
Pipher, J. L., \& Fazio, G. G. 2009, \apjs, 184, 18

\bibitem[Hansen et al.(2012)]{hansen12} Hansen, C.~E., Klein, 
R.~I., McKee, C.~F., \& Fisher, R.~T.\ 2012, \apj, 747, 22

\bibitem[Heitsch et al.(2008)]{heitschetal08}
Heitsch, F., Hartmann, L. W., Slyz, A. D., Devriendt, J. E. G.,
\& Burkert, A. 2008, ApJ, 674, 316

\bibitem[Hennemann et al.(2012)]{2012A&A...543L...3H} 
Hennemann, M., et al. 2012, \aap, 543, L3 

\bibitem[Hillenbrand(1997)]{hill97}
Hillenbrand, L.~A. 1997, \aj, 113, 1733

\bibitem[Hollenbach \& McKee(1989)]
{1989ApJ...342..306H} Hollenbach, D., \& McKee, C.~F.\ 1989, \apj, 342, 306

\bibitem[Jappsen et~al.(2005)]{jappsen}
Jappsen, A.-K., Klessen, R.~S., Larson, R.~B., Li, Y., \& Mac Low, M.-M. 2005, 
\aap, 435, 611

\bibitem[Kennicutt(1998)]{Kennicutt1998}
Kennicutt, Jr., R.~C. 1998, \araa, 36, 189

\bibitem[Kitsionas \& Whitworth(2002)]{kw}
Kitsionas, S., \& Whitworth, A.~P. 2002, \mnras, 330, 129

\bibitem[Klessen et~al.(1998)]{klessen}
Klessen, R.~S., Burkert, A., \& Bate, M.~R. 1998, \apjl, 501, L205

\bibitem[Kroupa(2002)]{kroupa}
Kroupa, P. 2002, Science, 295, 82

\bibitem[Krumholz et al.(2010)]{krumholz10} Krumholz, M.~R., 
Cunningham, A.~J., Klein, R.~I., \& McKee, C.~F.\ 2010, \apj, 713, 1120 

\bibitem[Krumholz et~al.(2007)]{krumholz}
Krumholz, M.~R., Klein, R.~I., \& McKee, C.~F. 2007, \apj, 656, 959

\bibitem[Krumholz et al.(2012)]{krumholz12} Krumholz, M.~R., 
Klein, R.~I., \& McKee, C.~F.\ 2012, ApJ, 754, 71

\bibitem[Krumholz \& McKee(2005)]{KM05}
Krumholz, M.~R., \& McKee, C.~F. 2005, \apj, 630, 250

\bibitem[Krumholz \& Tan(2007)]{KT07}
Krumholz, M.~R., \& Tan, J.~C. 2007, \apj, 654, 304

\bibitem[Lada \& Lada(2003)]{lada03} Lada, C.~J., \& Lada, E.~A.\ 2003, 
\araa, 41, 57 

\bibitem[Larson(2005)]{larson}
Larson, R.~B. 2005, \mnras, 359, 211

\bibitem[Li et~al.(2003)]{li}
Li, Y., Klessen, R.~S., \& Mac Low, M.-M. 2003, \apj, 592, 975


\bibitem[Martel et~al.(2006)]{mes06}
Martel, H., Evans, II, N.~J., \& Shapiro, P.~R. 2006, \apjs, 163, 122 (MES06)

\bibitem[McKee \& Ostriker(2007)]{mckee07} 
McKee, C.~F., \& Ostriker, E.~C.\ 2007, \araa, 45, 565 

\bibitem[Miller \& {Scalo}(1979)]{ms79}
Miller, G.~E., \& Scalo, J.~M. 1979, \apjs, 41, 513

\bibitem[Monaghan(1992)]{monaghan}
Monaghan, J.~J. 1992, \araa, 30, 543

\bibitem[Nenkova et~al.(2000)]{dusty}
Nenkova, M., Ivezi\'c, \v Z., \& Elitzur, M. 2000, Thermal Emission
  Spectroscopy and Analysis of Dust, Disks, and Regoliths, 196, 77


\bibitem[Offner et~al.(2009)]{offner}
Offner, S.~S.~R., Klein, R.~I., McKee, C.~F., \& Krumholz, M.~R. 2009,
  \apj, 703, 131

\bibitem[Ossenkopf \& Henning(1994)]{oh5}
Ossenkopf, V., \& Henning, T. 1994, \aap, 291, 943

\bibitem[Salpeter(1955)]{salpeter}
Salpeter, E.~E. 1955, \apj, 121, 161

\bibitem[Scalo(1986)]{scalo}
Scalo, J.~M. 1986, Fundamentals of Cosmic Physics, 11, 1

\bibitem[Shirley et~al.(2002)]{shirley}
Shirley, Y.~L., Evans, II, N.~J., \& Rawlings, J.~M.~C. 2002, \apj, 575, 337

\bibitem[Smith et~al.(2009)]{smith}
Smith, R.~J., Longmore, S., \& Bonnell, I. 2009, \mnras, 400, 1775

\bibitem[Stamatellos et al.(2011)]{2011MNRAS.413.1787S} Stamatellos, D., 
Maury, A., Whitworth, A., \& Andr{\'e}, P.\ 2011, \mnras, 413, 1787


\bibitem[Urban et~al.(2009)]{UrbanDG}
Urban, A., Evans, N.~J., \& Doty, S.~D. 2009, \apj, 698, 1341 (UED09)

\bibitem[Urban et~al.(2010)]{Urban10}
Urban, A., Martel, H., \& Evans, N.~J. 2010, \apj, 710, 1343 (UME10)

\bibitem[van der Tak \& van Dishoeck(2000)]{2000A&A...358L..79V} 
van der Tak, F.~F.~S., \& van Dishoeck, E.~F.\ 2000, \aap, 358, L79 

\bibitem[Vutisalchavakul \& Evans(2012)]{ve12}
Vutisalchavakul, N. \& Evans, II, N. J. 2012, submitted to ApJ

\bibitem[Williams et al.(2000)]{williams2000} Williams, J.~P., 
Blitz, L., \& McKee, C.~F.\ 2000, Protostars and Planets IV, 97 

\bibitem[Wu et~al.(2005)]{wu2005}
Wu, J., Evans, II, N.~J., Gao, Y., Solomon, P.~M., Shirley, Y.~L., \&
  Vanden Bout, P.~A. 2005, \apjl, 635, L173

\bibitem[Wu et~al.(2010)]{wu2010}
{Wu}, J., {Evans}, II, N.~J., {Shirley}, Y.~L., \& {Knez}, C.
  2010, \apjs, 188, 313

\bibitem[Wuchterl \& Tscharnuter(2003)]{wt}
Wuchterl, G., \& Tscharnuter, W.~M. 2003, \aap, 398, 1081


\bibitem[Young \& Evans(2005)]{chad}
Young, C.~H., \& Evans, II, N.~J. 2005, \apj, 627, 293

\bibitem[Young et~al.(2003)]{young}
Young, C.~H., Shirley, Y.~L., Evans, II, N.~J., \& Rawlings, J.~M.~C.
  2003, \apjs, 145, 111

\bibitem[Young et al.(2004)]{2004ApJ...614..252Y} Young, K.~E., Lee, J.-E., 
Evans, N.~J., II, Goldsmith, P.~F., \& Doty, S.~D.\ 2004, \apj, 614, 252

\end{thebibliography}
\end{document}